\documentclass[a4paper,12pt]{book}

\usepackage{a4wide}
\usepackage{amsmath,amsthm,amssymb}
\usepackage[latin1]{inputenc}
\usepackage{fancyhdr}
\usepackage{graphicx}
\usepackage{subfigure}
\usepackage{array}
\usepackage{hyperref}
\usepackage[german,english]{babel}

\usepackage{amsmath,amsfonts,amssymb}
\usepackage{latexsym}

\def \eref#1 {({\ref{#1}}) }

\def \nbb{\mathbb{N}}
\def \rbb{\mathbb{R}}

\def \zbb{\mathbb{Z}}

\def \xv{\vec x}
\def \yv{\vec y}

\def \<  {\langle}
\def \>  {\rangle}

\def \. { \,\! }

\def \cdotarg { \, \cdot \, }

\def \expltext#1 {\\ \text{\footnotesize{ (#1) }}\\}
\def \intercomm#1 {\\ \text{\footnotesize{ (#1) }}\\}
\def \undercomm#1 {\underset{\text{\scriptsize{ (#1) }}}}
\def \overcomm#1 {\overset{\text{\scriptsize{ (#1) }}}}

\def \tracenorm#1 { \| #1 \| _1 }
\def \hsnorm#1 { \| #1 \| _2 }

\def \vectorcomp#1 {
  \left( \begin{array}{c}
  #1
  \end{array}
  \right)  }

\def \pder#1#2 { \frac{ \partial #1 }{ \partial #2 }}

\newtheorem{defn}{Definition}
\newtheorem{lemm}[defn]{Lemma}
\newtheorem{prop}[defn]{Proposition}
\newtheorem{thm}[defn]{Theorem}
\newtheorem{corr}[defn]{Corollary}

\DeclareMathOperator{\vol}{vol}

\renewcommand{\xv}{ \boldsymbol{x} }
\renewcommand{\yv}{ \boldsymbol{y} }
\newcommand{\zv}{ \boldsymbol{z} }
\newcommand{\unitv}[1]{ \boldsymbol{e}_{(#1)} }

\newcommand{\expect}[1]{ \mathrm{E} \lbrack {#1} \rbrack}

\newcommand{\event}[1]{ \mathrm{\text{#1}} }
\newcommand{\conn}{ \event{CONN} }
\newcommand{\connpb}{ \event{CONN-PB} }
\newcommand{\discpb}[1]{ \event{{#1}-DISCONN-PB} }
\newcommand{\conndb}{ \event{CONN-DB} }
\newcommand{\discdb}[1]{ \event{{#1}-DISCONN-DB} }
\newcommand{\segm}[2]{ \event{SEGMENT-{#1}-{#2}} }

\newcommand{\intcube}{\int_{[0,1]^n}}

\newcommand{\muEqualT}[1]{ \mu^{T\mathrm{-eq}}_{#1} }

\newcommand{\longnlim}{ \xrightarrow{\;n \to \infty\;} }

\newcommand{\OmegaSp}{ \Omega_{\text{spatial}}}
\newcommand{\OmegaInt}{ \Omega_{\text{internal}}}

\newenvironment{quotepar}{
	\begin{quote}
}
{
	\end{quote}
}

\newcommand{\showurl}[1]{\href{#1}{#1}}

\newcommand{\unit}[2]{ #1\;\mathrm{#2}  }

\def\figext{eps}

\def \chapswitch {

\clearpage

\thispagestyle{plain}
\cleardoublepage
\pagestyle{fancy}

}

\newcommand{\specialChapter}[1]{

  \cleardoublepage
  \chapter*{{#1}}   
  \addcontentsline{toc}{chapter}{{#1}}
  \markboth{{#1}}{{#1}}

}

\newcommand{\specialChapterSilent}[2]{

  \cleardoublepage
  \specialChapterSilentLR{#1}{#2}
}

\newcommand{\specialChapterSilentLR}[2]{
  
  \clearpage
  \addcontentsline{toc}{chapter}{{#1}}
  \chead[\fancyplain{}{\slshape {#1}} ]{\fancyplain{} {\slshape {#1}} }
  \markboth{{#1}}{{#1}}

  {
    \renewcommand{\cleardoublepage}{\clearpage}
    #2
  }  

}

\fancypagestyle{plain}{%
\lhead[]{}
\chead[]{}
\rhead[]{}
\fancyfoot{}
\cfoot{-- {\thepage} --}

}

\numberwithin{defn}{chapter}
\numberwithin{equation}{chapter}
\begin{document}

\DeclareGraphicsExtensions{.jpg,.pdf,.mps,.png}

%\pdfoutline goto name{toc} count 0{Inhaltsverzeichnis}
%\pdfoutline goto name{notation} count 0{Notationskonventionen}
%\pdfoutline goto name{literatur} count 0{Literaturverzeichnis}

\thispagestyle{empty}
\begin{titlepage}

\vspace*{2cm}

\begin{center}

{\Huge \textbf{Statistische Analysen von Qualitätsmerkmalen mobiler Ad-hoc-Netze}}

\vspace*{1cm}

{\huge (Statistical Analysis of Quality Measures for Mobile Ad Hoc Networks)}

\vspace*{2.5cm}

{\Large  Abschlussarbeit \\ 
     \vspace*{0.2cm} im Studiengang Master of Computer Science \\
     \vspace*{0.2cm} der FernUniversität in Hagen }

\vspace*{3cm}

{\Large vorgelegt von \\ Henning Bostelmann \\ aus Soltau}

\vspace*{3cm}

{\Large Steinbach, März 2005}

\vfill

\end{center}

\end{titlepage}

\thispagestyle{empty}
\clearpage

%\end{document}

\thispagestyle{plain}
\cleardoublepage
\chead[\fancyplain{}{\slshape Contents} ]{\fancyplain{} {\slshape Contents} }

%\pdfdest name{toc} xyz

\tableofcontents

\clearpage
%\layout

\chead[\fancyplain{}{\slshape \leftmark} ]{\fancyplain{} {\slshape \rightmark}}

\thispagestyle{plain}

\chapswitch

\chapter{Introduction}

\section{Mobile ad hoc networks}

Mobile ad hoc networks (MANETs for short) are wireless networks which organize themselves.
They consist of a number of independent computing devices with wireless trans\-ceivers.
Not relying on a pre-existing infrastructure (such as access points or base stations),
these devices can exchange data with each other in a multi-hop fashion,
where each of the nodes is able to act as a router. 
A typical property of these system is that the position of network nodes,
and hence the network topology, is not predefined, but is set up at random
(or cannot be controlled)
and will even change dynamically if the nodes are mobile.

Such networks have been the subject of research since the beginning 1980s
(usually under the term \emph{packet radio networks}). But only 
recently, with the advent of commonly available, inexpensive wireless devices,
the subject has gained much attention and focus \cite{CCL:imperatives}.

While MANETs are not currently in widespread use,
there are a number of promising applications:
Traffic information might be transmitted to cars on a motorway
via an ad hoc network composed of on board communication systems \cite{HBE:fleetnet}.
Pedestrians carrying mobile phones or PDAs might use a MANET for mobile
data access, and firefighters could use such networks to gather 
critical information on scene \cite{JCH:siren}.
MANETs could prove to be particularly useful in situations where 
no network infrastructure is available, such as in 
disaster areas. 
Another interesting use case is their ability to extend an existing infrastructure,
such as WLAN hotspots, beyond the range of the installed access points.

A special case of MANETs, which is specifically the target of current research,
are \emph{sensor networks} \cite{ASS:sensors}
which are formed of small, inexpensive, autonomous devices that 
are able to capture measurement data of some kind and transport them to a central
location using ad hoc network techniques. Such sensor nodes are usually not mobile,
but might be deployed in an area at random, e.g. by dropping them from a plane.
Sensor networks might deliver valuable data for use in environmental monitoring,
agriculture, or forest fire detection, just to name some examples.

On the technical side, MANETs would typically be based on existing wireless
technology, most notably IEEE 802.11 WLAN \cite{XS:wlan} and Bluetooth \cite{KRSW:bluetooth}.
While MANETs thus inherit a number of difficulties that are inevitably connected to
wireless communication -- such as low reliability of links
and the hidden-station problem --, there are a number of characteristics and issues
that are specific to the ad hoc domain:

Due to the lack of a central controller, virtually all algorithms 
used on layer 3 and beyond
must be distributed to the network nodes in order to provide for scalability and 
fault-tolerance. This applies not only to the application layer, but in particular
to the routing protocols used. These are especially important since the network
topology may change frequently at run time, so that traditional routing approaches 
cannot be applied as usual. A~number of specialized routing protocols have been 
developed that are optimized for the specific needs of MANETs \cite{Raj:routing}.

Further, when using mobile or sensor devices, one is concerned with the problem of energy
consumption: The network nodes are usually powered by batteries whose capacity 
may restrict the online time of each node (since users can only recharge their devices
at intervals) or, in the case of maintenance-free sensor devices, may even 
limit the lifetime of nodes. A good part of the power consumption
of the nodes is in fact due to their radio transceiver. 
Usual power-saving strategies are based on switching the devices into some inactive mode
when not used; this is not favourable in ad hoc networks, however, since 
each device may be needed as a router. Thus, the \emph{range assignment problem}
is of central importance for MANETs: If the transmitting power (hence the radio range) 
of the network nodes is chosen too large, this amounts to a waste of battery resource.
(It might also put limits to spatial channel reuse.) On the other hand, choosing
the radio range too small may impact the network quality,
since the number of point-to-point links is reduced. Therefore, a critical
point in MANET design is to select the right radio range for a given 
density of nodes (or vice versa). Also, methods have been investigated to choose the 
radio range of each node dynamically \cite{KKKP:power_consumption,RR:topology_control}.

Last but not least, a large number of research activities focus on the 
development of applications
using MANETs, on specialized middleware (e.g. \cite{Rot:middleware,Her:middleware}), 
and on security aspects \cite{BH:security_session}.

\enlargethispage{0.5cm}

In evaluating design proposals for MANET systems, it is usually very hard to
actually verify them in real measurements: Such experiments need to rely
on a prototype implementation, which is usually available only very late
in the development cycle; moreover, they are quite cost-intensive, considering
the large number of network nodes involved. Due to these obstacles,
only quite few experimental evaluations have been performed \cite{MBJ:quantitative_lessons, KNG:experimental_eval},
with the MANET size being far below 100 nodes -- which 
is rather on the low end for possible applications,
considering that sensor networks of several 10.000 nodes are being discussed.
Evaluation of protocols, etc. are therefore often based on numerical simulations:
Using a statistical model that describes the spatial distribution of nodes 
and, for models with mobility, their movement on the deployment region, it is possible 
to evaluate the performance of routing algorithms, or to 
analyse general quantitative properties of MANETs; several off-the-shelf
network simulators are available.\footnote{
  Examples include OPNET (\showurl{http://www.opnet.com/products/modeler/home.html}),
  NS-2 (\showurl{http://www.isi.edu/nsnam/ns/}),
  and GloMoSim (\showurl{http://pcl.cs.ucla.edu/projects/glomosim/}).
}
On the statistical side, a variety of different mathematical models are used
to describe the mobility of nodes \cite{CBD:mobility_models}, 
one of the best-known being the \emph{random waypoint model}.
Recently, a number of publications have questioned the accuracy 
of the results of such simulations, both with respect
to the properties 
of the statistical model involved \cite{CSS:accuracy_simulators,YLN:random_waypoint_harmful}
and to over-idealized modelling assumptions \cite{KNG:experimental_eval};
the accuracy of simulations must therefore be regarded as an open issue.
Exactly solvable models, or other analytical results in MANET models,
are currently only very rare, largely due to the complexity of the problems involved
(see, however, the next section).

\section{Quality and connectedness}

Let us take a closer look at the results known in the literature for
measuring the quality of MANETs. Here we do not refer to the performance
of routing algorithms or higher-level protocols, but we are rather interested
in restrictions on the lower layers, related e.g. to inter-node connectivity.

One natural question in this context is whether the MANET is connected,\footnote{
Sometimes, the term \emph{strongly connected} is used to describe this situation.}
i.e. whether there is a multi-hop network path  
between each pair of nodes in the MANET. Since we are dealing with nodes
which are distributed at random, we are thus asking for the \emph{probability}
that the network is connected.

Some early works \cite{PPT:packet_radio_connectivity,Pir:radio_connectivity}
established asymptotic estimates for the probability of connectedness
in 1-dimensional systems and conjectured analogue results for 2-dimensional systems.
Here the nodes were distributed on an area (or line segment) according
to a Poisson process of homogeneous density. The authors dealt with the probability
that the area is completely covered by the MANET, i.e. that each point
is in the range of at least one network node. 
Recently, the results for 2-dimensional systems were made precise by
Xue and Kumar \cite{XK:neighbours}.
Denoting the number of network nodes by $n$, these results show that
the mean local density of network nodes must to grow 
by a factor $\Theta(\ln n)$ in the limit $n \to \infty$
if one wants to keep connectedness (or coverage of the area).

The probability of connectedness was also considered
by Santi and Blough \cite{SanBlo:transmitting_range}, based on earlier similar work
\cite{SanBlo:connectivity,SBV:range_assignment}. The authors 
derive asymptotic estimates mainly for the 1-dimensional system and present
numerical (simulation) results also for 2- and 3-dimensional systems.

Bettstetter \cite{Bet:minimum_node_degree} generalized this to the 
even stronger condition of $k$-connectedness. (The network is called
\emph{$k$-connected} if between each node pair, there are at least $k$ independent
network paths.) Using results from the theory of random graphs \cite{Pen:k_connect},
he established analytical estimates for the 2-dimensional case and verified 
them with numerical results. He also calculated the probability that 
none of the nodes is completely isolated in the network.

More general quality measures have been defined by Roth \cite{Rot:critical_mass}
and investigated in a numerical simulation. Here, not the probability of connectedness
is taken as a quality indicator (since, as the author notes, 
connectedness is a rather strong condition for MANETs); instead,
measures based on the number of separated network segments, the size of these segments,
and the dependence of these on changes in the network (e.g. a node being switched off), 
are being considered.

Further, an analytical estimate of the bandwidth available to each node
has been given by Gupta and Kumar \cite{GK:capacity}.
The authors show that this bandwidth is of the order
$W / \sqrt{n \log n}$, where $W$ is the bandwidth of the point-to-point links;
thus, the throughput that each node is able to use rapidly decreases with the network size.

\section{Scope of this work}

In this work, we shall be concerned with the evaluation of quality measures
for ad hoc networks described by statistical models. 
As in the last section, we will not consider complex MANET protocols,
but rather focus on simple models for connectivity between the network nodes;
we shall establish explicit analytical results for the expected quality
of MANETs on this level.

We will set out from a statistical model of MANETs, similar to that
considered in \cite{SanBlo:transmitting_range}, where a number of nodes 
is distributed independently at random in a given area. Focusing
on the 1-dimensional situation (which might be interpreted as a network
of cars on a road, or pedestrians on the sidewalk), we will show that the
model is exactly solvable, and derive precise results for the 
probability of connectedness and other quality measures. Comparing these
results to the literature, we will find that the numerical results
both by Santi et al.~\cite{SBV:range_assignment} and by Roth~\cite{Rot:critical_mass}
can be explained by our calculations, although they were based on different
modelling assumptions.

The work is organized as follows:

In Chap.~\ref{ModelChap}, we will define the general framework of statistical modelling
that our calculations are based on. Idealizations and assumptions 
involved in this modelling will be discussed.

Chapter~\ref{ConnectChap} focuses on a specific model, the 1-dimensional MANET
with homogeneously distributed nodes. Neglecting boundary effects, we will explicitly
calculate the probability of connectedness for a fixed MANET size, and 
establish an asymptotic formula for the limit of large MANETs.
This allows us to compare our results to existing work, in particular \cite{SanBlo:transmitting_range}.

More general quality measures will be defined and analysed in Chap.~\ref{QualityChap}.
We classify these quality measures according to their scaling behaviour.
Using the same model as in Chap.~\ref{ConnectChap}, we are able to establish 
explicit values for these measures in the 1-dimensional case, both at
fixed size and in the limit of large systems. We compare
our results to the numerical data from the literature~\cite{Rot:critical_mass}
and discuss similarities and differences.

Chapter~\ref{MiscChap} discusses extensions of the results established 
in the previous chapters to more general situations. As an example,
we explicitly treat a model where network nodes are switched off at random.
An outlook is given to results in higher-dimensional systems and other 
extensions of the current results.

Two appendices cover matters that are somewhat outside the main line of argument:
Appendix~\ref{MathApp} develops some mathematical results used in the main text,
while Appendix~\ref{SantiApp} discusses certain issues found in the comparison
of our results with~\cite{SanBlo:transmitting_range}. The reader is also referred
to the index of notation on page~\pageref{NotationApp}.

\chapswitch

\chapter{Statistical Models for Ad Hoc Networks} \label{ModelChap}

It is a characteristic property of ad hoc networks that,
unlike in traditional
infrastructure-based networks, the positions
of the network nodes cannot be controlled.
Therefore, it is generally useful to assume
that the nodes are distributed at random in some area -- in 
particular when the number of nodes is large --,
and to use methods from probability theory in order
to analyse the behaviour of the system.

Such an analysis can, quite generally,
be divided into two steps:
\begin{enumerate}
  \renewcommand{\theenumi}{(\roman{enumi})}
  \renewcommand{\labelenumi}{\theenumi}
  \item the definition of a mathematical model that represents the 
  	 situation under discussion,
  \item the evaluation of this model and an analysis of its predictions.
\end{enumerate}

Both steps will, in general, involve simplifications and approximations
of the ``exact'' situation: In the definition of the model,
one decides on which aspects of the real situation should be modelled 
and which should be omitted; in the evaluation of the model,
one often uses approximations (such as asymptotic expansions or limits)
that only provide a certain level of precision. 

Certainly, the two steps are not independent: When defining the mathematical
setup, one naturally has to take care not to define the model too detailed,
in order to keep the complexity of evaluation within reasonable limits.
So there usually is a trade-off between precision in modelling and precision
in evaluation.  

However, it seems important to keep the distinction between the two steps clear,
and to define clear interfaces between them. 
This is particularly important for the comparison between different models
or numerical approximations. It has recently been 
exposed \cite{CSS:accuracy_simulators} that the different simulation approaches
can lead to very different results in the evaluation of protocols, even 
with regard to qualitative predictions. 
In this kind of situation,
it would certainly be helpful to have a clear distinction between modelling
and evaluation, since this might serve to clarify differences between the approaches,
and to determine whether the difference lies in numerical approximation or in 
the general assumptions.
It is also possible that an ``informally'' defined mathematical model (that is 
defined only by specifying a numerical approximation) might include implicit 
properties that were not meant to be included in that way, as has recently
been discovered with regard to the spatial node distribution in the random
waypoint model~\cite{YLN:random_waypoint_harmful}.

In this chapter, we will describe a general framework for 
the statistical description of MANETs, and discuss the 
assumptions and simplifications associated with it.
We will try to define this framework quite generally,
although only a very specific case will be analysed in 
detail in later chapters. This done is to provide a broader
discussion of modelling assumptions and to hint at 
extension options for more complex systems.

The formalism that describes the statistical behaviour of nodes
is presented in Sec.~\ref{ProbModelSec}, while general assumptions
related to the network model are discussed in Sec.~\ref{RandomVarSec}.
The evaluation of specific models is
part of Chapters~\ref{ConnectChap} to~\ref{MiscChap}.

\section{A general statistical model} \label{ProbModelSec}

We analyse an ad hoc network of $n$ network nodes. These nodes are,
at fixed time, distributed at
random over some volume or area.

First, we will define the statistical side of the model, i.e.
define the random location of nodes and, optionally, 
additional inner parameters.
We will assume a sample space of the following form:
\begin{equation} \label{OmegaDef}
  \Omega_n = (\OmegaSp \times \OmegaInt)^n.
\end{equation}
Here $\OmegaSp$ denotes the sample space which describes the location 
of a single network node. It would usually be a subset of $\rbb^d$, where
$d \in \{1,2,3\}$.  The sample space $\OmegaInt$ describes internal 
parameters of the node; e.g. the node might be switched off at a certain
probability, or its transmission range might vary according to a random
process. We are describing each of the $n$ network nodes with the same 
sample space; note that this does \emph{not} yet imply that the corresponding
probability distribution is equal for each node, or independent between nodes.

Another part of the sample space might be used to describe global
random features of the model, e.g. the position of a shielding wall that disturbs
network transmission. However, we will not make use of such an alternative here.

On $\Omega_n$, we then need a probability measure $\mu_n$ 
which describes the distribution
of nodes. We will specify further assumptions on $\mu_n$ below.

Let us now discuss the general modelling assumptions that are already 
implicitly included in the definition \eqref{OmegaDef}, and additional
assumptions that are typically made in order to simplify the discussion.

\paragraph{Fixed number of nodes.}

With the above setup, we have assumed that the number of nodes in the network
is fixed, i.e. it does not vary at random. This is certainly a restriction,
since in a real scenario, the number of nodes might not be
determined \emph{a priori}; e.g. users might enter or leave the range of the network, 
or switch their 
devices off in order to save power. On the other hand, while it is possible
to include a varying number of nodes into the setup of the sample space,
this would increase the mathematical complexity of the system considerably, 
since it would require to move from a usually finite-dimensional space or manifold
$\Omega_n$ to an infinite-dimensional situation. 

In our context, it seems to be justified to stay with the situation of fixed $n$
for two reasons: First, when considering a large number $n$ of nodes, one
expects that the effect of a varying number of nodes is small, and 
that it suffices to take only the mean number into account. Second,
if we explicitly need to account for nodes dynamically joining the network,
we can always model them as nodes which are randomly switched off
from the network, including this aspect as a feature of $\OmegaInt$. 
Such an analysis will be presented in Section~\ref{VaryingNodeNumberSec}.

Note that while we model the statistical situation only for fixed $n$,
we will usually be interested in the expectation values of random variables
``for large $n$,'' i.e. in the limit $n \to \infty$.

\paragraph{Static situation.}

Our analysis restricts to the situation of the network at fixed time. 
This may seem to be a bit contrary to our goal to describe \emph{mobile}
ad hoc networks. However, mobility of the nodes does not imply that
the \emph{probability} to find a node within a specific region varies with time.
In fact, since the movement of nodes (e.g. of visitors
in a shopping centre) will usually not be under our control in realistic situations, 
the best assumption might be that at any fixed time, nodes are
distributed according to a static probability distribution.

In other approaches to the statistical description of MANETs, one
often introduces an explicit model for the random movement of nodes
(such as the random waypoint model). However, even in these models, one would
assume that the spatial distribution of nodes stays constants over time,
or rather consider it as a problem in the model 
if this is not the case \cite{YLN:random_waypoint_harmful}.
In fact, network simulators may need a ``warm-up phase'' until 
a ``steady state'' in spatial distribution is reached.

Regarding time averages of random variables, we can deduce results
from our static model if the network system is \emph{ergodic:}
This means that time averages can be replaced with averages over the 
spatial coordinates of nodes at fixed time, which we can handle directly.
Equivalently, we may require that for each initial configuration of the network,
we reach almost every other possible configuration 
after waiting for a sufficiently long time 
(\emph{Birkhoff's ergodic theorem;} see \cite{Pet:ergodic_theory}).
Ergodicity of the system is not guaranteed and depends on a 
mobility model still to be chosen; however, lacking
specific information on the time dependence of the system,
it seems to be a natural assumption for our purposes.

Certainly, our model could easily be extended to describe non-static
situations by making the probability measure dependent on the
time $t$.
However, our setup generally does not allow to describe aspects of the system
that involve direct time dependence of random variables.
For example, assuming ergodicity, we might be able to answer the question: 
``For what portion of the time is a specific node connected to the network?'',
but our setup does not allow to discuss the question:
``For how long does a specific node stay connected to the network, once
it has established a connection?'' In our discussion of quality measures,
such time dependencies will not be relevant; for a discussion of routing
algorithms, on the other hand, they may be a crucial feature.

\paragraph{Independence.}
In addition to the above assumptions, we will apply another 
simplification, namely the statistical independence of the nodes. On the mathematical
side, this means that our probability measure is reduced to a product
\begin{equation}
  \mu_n = \prod_{j=1}^n \mu^{\text{node}}_{j},
\end{equation}
where $\mu^{\text{node}}_{j}$ are probability measures on $\OmegaSp \times \OmegaInt$,
describing the distribution of a single node. 

On the modelling side, this means that the different nodes
will have no mutual influence on their positions (or other internal states). 
This need not be fulfilled in realistic situations;
for example, in a traffic jam, the position of a specific car 
will very well be influenced by the position of the car in front of it.
Such aspects cannot be described when making the above assumption;
however, it seems plausible that in many situations, such effects
will not play a major role.

\paragraph{Identical distribution.}

In addition to the independence of nodes, we will assume in all our examples
that the node are distributed \emph{identically}, i.e. that all
$\mu^{\text{node}}_{j}$ are in fact equal:
\begin{equation}
	\forall j: \; \mu^{\text{node}}_{j} = \mu^{\text{node}}_{(0)}
\end{equation}
This seems to be a natural assumption if there is only one type of
node involved in the network. It does not cover a situation where 
certain nodes are distinguished from others, e.g. where certain users
prefer a specific part of the area. We might, however, still cover
these situations when modelling this behaviour within $\OmegaInt$; 
that is, we would let the user choose ``at random'' which area he prefers,
while preserving the identical distribution. (However, such models 
will not be covered in this text.)

\paragraph{No feedback.}

In all of the following text, we will assume that the probability 
measure $\mu_n$ is given \emph{a priori} as a fixed quantity,
and that it does not depend on the details of the MANET quality;
alternatively speaking, there is no ``feedback'' from the random
variables to the probability distribution.

Of course, one might in principle think of a situation in which users prefer
to visit areas where the MANET quality is usually good, or in which 
they tend to switch off their devices if they loose connectivity 
for an extended period. Such aspects would need to be modelled in form
of a (supposedly complicated) relation in $\mu_n$, e.g. a differential 
equation, that would leave us with the task of finding a solution 
for $\mu_n$ before calculating expectation values. However,
this lies far beyond the scope of the current presentation.

\section{Random variables} \label{RandomVarSec}

Having specified the statistical behaviour of the system, we will
now turn to a description of the random variables. Random variables
would include, e.g., the number of network segments, the number of
nodes that a specific node is connected to, or the spatial distance between two nodes.
In the general mathematical setting, a random variable is an (integrable)
function
\begin{equation}
	F: \Omega_n \to \rbb.
\end{equation}
As usual, we consider the \emph{expectation value} of $F$, defined as
\begin{equation} \label{expectDef}
	\expect{F} :=  \int_{\Omega_n} d \mu_n(\omega) F(\omega).
\end{equation}
and interpreted as the statistical mean of $F$. We sometimes write it as
$\overline{F}$ for short.

Without putting too much emphasis on the mathematical formalism, 
it should be noted that a random variable itself does not 
include the statistical description of the model; for
each fixed $\omega \in \Omega$, its value $F(\omega)$ is simply
the ``deterministic'' value of the function in the elementary event $\omega$.
The statistical behaviour is described via the expectation value alone.

Usually, the definition of random variables will depend on $n$, as well as on 
other parameters of the system (such as the range $r$ of the radio devices). 
As above, we will often not denote this dependence explicitly, in order not
to overburden the notation; in case were it becomes necessary,
we will explicitly write $F^{(n)}$,  $F^{(n,r)}$, or similar.

In order to fix our notation, let us briefly introduce some special random variables,
which are connected to \emph{events} on $\Omega_n$. An event $\event{EV}$ is
a subset of the sample space $\Omega_n$; as an example, take the 
event $\conn$ which contains all points of $\Omega_n$ that correspond to situations
where the MANET is strongly connected.
For notational purposes, we will often write events as $M_\event{EV} \subset \Omega_n$
when referring to it as a set. To each such event corresponds its 
\emph {characteristic function} $\chi_\event{EV}$, defined as
\begin{equation} \label{charactDef}
	\chi_\event{EV} (\omega) = \begin{cases}
		1 & \text{ if } \omega \in M_\event{EV}, \\
		0 & \text{ otherwise},
	\end{cases}
\end{equation}
which is a random variable in our sense. Its expectation value
\begin{equation} \label{probabDef}
	P_\event{EV} := \expect{ \chi_\event{EV} }
\end{equation}
is the \emph{probability} that the event $\event{EV}$ will occur.

After these formalities, let us discuss our modelling assumptions
on the random variable side more closely. It is difficult however
to investigate properties of specific random variables
(such as connectivity of the network) without specifying a 
concrete model, which we postpone to Chap.~\ref{ConnectChap}.
However, we shall discuss a number of general assumptions
on the random variables, and how we wish to handle them.
This follows a recent discussion by Kotz et~al. \cite{KNG:experimental_eval}
who identified a number of common assumptions in MANET models and compared
them with experimental results. The authors criticized these assumptions
as begin too restrictive for realistic scenarios; we will 
in fact stick to all of these assumptions in this text, and 
will argue in the following why they are justified in our simple 
situation.

\paragraph{The world is flat.}

While radio propagation is a 3-dimensional phenomenon, the nodes of a MANET
are usually distributed over some 2-dimensional (e.g. 
sensors deployed in an area) or even 1-dimensional region (e.g. cars on a road).
Truly 3-dimensional situations will only very seldomly be found in practice,
since ceilings in buildings, etc. usually block radio propagation.
If some network nodes are located in vertically exposed positions (e.g. on hills),
it seems more appropriate to include this effect in the model by modifying
their individual radio range (see below) rather than turning to a 3-dimensional
description of their position. In most of this work, we shall restrict 
to the 1-dimensional case for simplicity.

\paragraph{A radio's transmission area is circular.}
2-dimensional MANET models usually assume that the range of network nodes is not dependent
on direction. While this seems to be a natural assumption at first, 
it is often not realized in experiment \cite[Fig.~1]{KNG:experimental_eval};
in particular, commonly used antennas are not omnidirectional.
However, for the 1-dimensional situation that we will consider, 
these properties will obviously be of less importance.

\paragraph{Signal strength is a simple function of distance.}

In generalization of the last point, it may even be
difficult in experiment to find any simple relation between the
spatial distance of nodes and the signal quality on point-to-point links,
since the signal strength is influenced by radio reflection, shielding obstacles
(including e.g. the person carrying a mobile device), and other 
effects that are difficult to control. In fact, the data presented in 
\cite{KNG:experimental_eval} suggests that the radio range of nodes 
should rather be described by a statistical process.
We might include this behaviour in our model by assigning the radio range of
nodes at random, albeit at the cost of
a much increased complexity in evaluation.
However, for the simple connectivity properties we will consider, it seems reasonable
that only the \emph{mean} radio range of nodes will be relevant for our results
-- see also the discussion in Sec.~\ref{outlookSec}.

\paragraph{All radios have equal range.}

We will assume in our specific models that all nodes are equal with respect to their
radio range. Due to varying background noise, differences in device configuration,
and also for reasons named above, this may not be given in experimental situations.
Again, we might include this in our model by considering the radio range of
nodes as a random variable, or choosing it dependent on the node's spatial position;
we will however refrain from doing so in the present work.

\paragraph{If I can hear you, you can hear me (symmetry).}

Many MANET protocols discussed in the literature rely on network links
to be bidirectional, while it has been stated in \cite{KNG:experimental_eval}
that this assumption is often not valid in practice; in particular,
packet collisions may lead to unidirectional links. 
While this may be a crucial feature for routing protocols, 
unidirectional links should not affect our simple evaluations of 
connectivity: We aim at a description of the connectedness
between nodes and disregard packet loss rates, etc.

\paragraph{If I can hear you at all, I can hear you perfectly.}

For our evaluation of MANET connectivity, we will focus on the question
whether a point-to-point link between two nodes can be established,
and will not aim at a calculation of network throughput, packet loss,
or error rates. Therefore, we can assign a sharply defined ``range''
to each node, below which we assume point-to-point links to be established,
and beyond which no communication is possible. 
Certainly, for a more detailed analysis of MANETs, it should be taken
into account that there is no sharp spatial cutoff for connectivity,
but that the signal strength decays gradually with increased distance
from a node.

\chapswitch
\chapter{The Connectivity of 1-dimensional Networks} \label{ConnectChap}

We will now proceed to a specific MANET model
which we will analyse in detail. This model
restricts to the 1-dimensional situation, 
i.e. the nodes are deployed at random along a straight line.
One might think here of pedestrians moving along
sidewalks, or of cars on a road, that carry wireless
devices. While this model is relevant at least
for parts of the proposed applications, it turns out to be particularly
simple in mathematical description, so that we can
derive explicit analytical results e.g. for the probability
of connectedness.
 
In Sec.~\ref{modeldefSec}, we will first give a definition 
of the model and introduce specific assumptions.
Then, in Sec.~\ref{periodicSec} we consider a variant of the model -- the model
with \emph{periodic boundary conditions} -- which 
allows us to calculate the probability of connectedness,
both for a fixed node number $n$ and in the limit 
$n \to \infty$. Section~\ref{disconnectedSec} will then
return to the model with ``usual'' boundary conditions,
transfer our results to that situation, and compare the
outcome with analytical and numerical results known in the literature.

\section{Definition of the model} \label{modeldefSec}

As mentioned in the introduction, we will now consider a 
1-dimensional system with $n$ network nodes distributed at random.
More specifically, we assume that the network nodes are distributed
on an interval $[0,\ell]$, that is, we set
\begin{equation} \label{OmegaNOneD}
   \OmegaSp = [0,\ell],\quad
   \Omega_n = [0,\ell]^n,
\end{equation}
where the space $\OmegaInt$ is trivial, i.e. we consider no
additional internal random parameters of the network nodes.
For simplicity, we will assume that the $n$ nodes are
distributed identically and independently, according to
the \emph{equal distribution} on $\OmegaSp$. This means that
\begin{equation} 
   d\mu^{\text{node}}_{(0)} = \ell^{-1} dx,
   \quad
   d\mu_{n} = \ell^{-n} d^nx,
\end{equation}
This fixes the statistical behaviour of the system. It still remains
to define the random variables of interest. 

In this chapter, we will mainly be
concerned with the question whether the network is \emph{connected,}
i.e. whether all nodes are able to communicate with each other
in a multi-hop fashion. To that end, we will assume that all nodes
have a fixed (and identical) radio range of $r$. Two nodes 
with coordinates $x_i$ and $x_j$ can 
communicate directly with each other if 
\begin{equation} \label{directComm}
   | x_i - x_j | < r.
\end{equation}
It is then clear what ``connectedness'' means in the model.
This is essentially the situation considered by Santi and 
Blough~\cite{SanBlo:transmitting_range}, who derived estimates
on the probability of connectedness in the limit $\ell \to \infty$. 

Before we proceed to the precise definition of random variables
and the calculation of expectation values, let us first 
discuss some general properties of the random variables 
involved, since the model has some symmetries that
we will exploit to ease our calculation later on.

The first of these properties relates to the fact that 
the behaviour of the system does not depend on $\ell$ and
$r$ explicitly, but that is stays the same when 
$r$, $\ell$ and all coordinates are scaled by a common
factor (there
is no ``fixed length scale'' in the system). Heuristically,
this is easily understood directly from the model;
since we will make a lot of use of this property, let us
however describe it more formally.

\begin{defn} \label{scalingDef}
\sloppy
In the 1-dimensional MANET model,
a family of random variables $F^{(n,\ell,r)}: [0,\ell]^n \to \rbb$
is called \emph{scaling} if
\begin{equation*}
   \forall n \in \nbb \;\;
   \forall \lambda > 0 \;\;
   \forall \xv \in [0,\ell]^n:
   \quad
   F^{(n,\ell,r)}(\xv)  = F^{(n,\lambda \ell, \lambda r)}(\lambda \xv).
\end{equation*}
\fussy
\end{defn}

In fact, all random variables considered in the following will be scaling;
this is due to the fact that the connectivity between two nodes
is not affected by scaling, cf. Eq.~\eqref{directComm}. The consequence
of this property is that expectation values are indeed 
dependent on the ratio $r/\ell$ only:

\begin{prop}
Let $F^{(n,\ell,r)}$ be a scaling family of random variables.
Then we have for all $r > 0$, $\ell > 0$:
\begin{equation*}
   \expect{F^{(n,\ell,r)}}  =    \expect{F^{(n,1, r/\ell)}}.
\end{equation*}
\end{prop}
\begin{proof}
By definition of the expectation value, we have
\begin{equation}
	\expect{F^{(n,\ell,r)}}
	= \int_{[0,\ell]^n} d^nx \, \ell^{-n} \,	F^{(n,\ell,r)} (\xv)
	= \ell^{-n} \int_{[0,\ell]^n} d^nx	\, F^{(n,1,r/\ell)} (\ell^{-1} \xv),
\end{equation}
using the scaling property with $\lambda = \ell^{-1}$. 
Now a simple substitution of variables $x_i' = \ell^{-1} x_i$
leads us to
\begin{equation}
	\expect{F^{(n,\ell,r)}}
% \ell^{-n} \int_{[0,\ell]^n} d^nx	\, F^{(n,1,r/\ell)} \big(\frac{\xv}{\ell}\big)
	= \int_{[0,1]^n} d^n x'	\, F^{(n,1,r/\ell)} ( \xv')
	= \expect{F^{(n,1,r/\ell)}},
\end{equation}
as proposed.
\end{proof}

Since in what follows, our results will relate to expectation values
or probabilities, they will therefore depend on $n$ and $r/\ell$ only.
Alternatively speaking, we can refer to the sample space $\Omega_n = [0,1]^n$
at any scale and use the ``normalized radio range'' $\rho := r/\ell$ in place of $r$,
thus reducing the number of parameters by one.
Since all our random variables will be scaling, it is justified to use
this simplified model only; we will return to the explicit parameters $\ell$
and $r$ only for comparison with experiment or other publications.

The next general property is related to the fact 
that all $n$ nodes are treated as equal in the model.

\begin{defn}
In the 1-dimensional MANET model, 
a random variable $F$ is called \emph{symmetric} if, for any
permutation $\sigma : \{1,\ldots,n\} \to \{1,\ldots,n\}$, it holds that
\begin{equation*}
  \forall \xv \in \Omega_n: \;
  F(x_1,\ldots,x_n) = F(x_{\sigma(1)},\ldots,x_{\sigma(n)}). 
\end{equation*}
\end{defn}

All random variables we consider in the following will be
symmetric.\footnote{
In fact, 
for any random variable $F$ we might always define the
symmetric random variable
\begin{displaymath}
  F_\text{Symm} (\xv) := \frac{1}{n!} \sum_\sigma F(x_{\sigma(1)},\ldots,x_{\sigma(n)})
\end{displaymath}
due to the symmetry of the underlying integration measure, 
we easily find $\expect{F}=\expect{F_\text{Symm}}$.
}
This property has the consequence that we can calculate 
expectation values more easily: In the integral
\begin{equation}
   \expect{F}  = \intcube d^nx F(\xv),
\end{equation}
we can split the integration region $[0,1]^n$ into $n!$
regions where the coordinate values are sorted in a specific order,
i.e. 
$R_1 = \{x \,|\, x_1 < x_2 < \ldots < x_n\}$,
$R_2 = \{x \,|\, x_2 < x_1 < x_3 < \ldots < x_n\}$ etc., ignoring
sets of volume zero. Since all these regions have identical volume,
and since a symmetric random variable $F$ is not affected by a
change in the order of variables, one has in this case
\begin{equation}
   \expect{F}  = n! \int_{R_1} d^nx F(\xv).
\end{equation}
More explicitly, we can express this as
\begin{equation} \label{sortedInt}
   \expect{F}  = n! \int_0^1 dx_1 \int_{x_1}^1 dx_2 \ldots \int_{x_{n-1}}^1 dx_n F(\xv);
\end{equation}
this form is often convenient, whenever $F$ can be 
formulated easier in the ``sorted'' coordinates $x_1 \leq \ldots \leq x_n$.

\section{Connectivity with periodic boundary conditions} \label{periodicSec}

Up to now, the system we defined was identical to the one
considered by Santi and Blough~\cite{SanBlo:transmitting_range}.
In this model, one would define that in sorted coordinates
$x_1 \leq \ldots \leq x_n$, the node $i$ is connected to its
neighbour $i+1$ if $x_{i+1}-x_i<\rho$; for the nodes $1$ and $n$, 
however, there is no left-side or right-side neighbour, respectively,
which they could connect to.
While this definition seems somewhat natural, it leads to an increased complexity
if one wants to derive analytical results: It includes a description
of the effects at the boundary of the network, which one implicitly 
has to account for in any calculations. 

As a method to overcome these difficulties, we will introduce 
\emph{periodic boundary conditions} in our model: We will say that
the left-most node is connected to the right-most one if
\begin{equation}
 x_1+1 - x_n < \rho.
\end{equation}
This amounts to a periodic extension of the node coordinates to 
the region outside $[0,1]$. One might also think of the nodes
being located on a closed path rather than an interval.\footnote{
The use of periodic boundary conditions is a well-known technique
for dealing with similar types of boundary problems; it has also
been applied the analysis of MANET connectivity before \cite{Bet:minimum_node_degree}.
} 

While this change seems to be a bit technical, it is justified for two
reasons: First, 
we are interested in the behaviour of the MANET in the ``bulk''
and not at the boundaries; it is thus reasonable to eliminate
boundary effects via the periodic extension.
(In fact, in a realistic scenario such as the shopping center
example considered by Roth \cite{Rot:critical_mass}, the paths
that users are located on would include both closed curves and 
open segments, and thus a ``disconnected'' boundary condition is 
\emph{a priori}
not more realistic than a periodic one.)
Second, and more importantly, it is expected that in the limit
of large MANETs ($n \to \infty$), these boundary effects play no rôle,
and both models lead to the same results. We will explicitly
show this for the probability of connectedness in Sec.~\ref{disconnEstimatesSec}.

\subsection{Transformation of the probability space} \label{transformSec}

On the analytical side, the introduction of periodic boundary conditions
amounts to a change in the random variables (the probability distribution
is unchanged); it results in the following property.

\begin{defn}
A random variable $F:[0,1]^n \to \rbb$ is called \emph{translation invariant}
if\footnote{
Note that the definition does \emph{not} refer to sorted coordinates.
}
\begin{equation*}
 \forall \xv \in [0,1]^n \;\;
 \forall \lambda \in \rbb: \quad
 F(x_1, \ldots, x_n) = F(x_1 + \lambda, \ldots, x_n + \lambda),
\end{equation*}
where the function $F$ is taken to be periodically continued to $\rbb^n$,
i.e. $F(x_1+1,x_2,\ldots) = F(x_1,x_2,\ldots)$ etc.
\end{defn}

We will later see why all relevant variables in our context are in fact
translation invariant. Let us first analyse the consequences of
this property. To that end, let $F$ be a symmetric and translation
invariant (as well as scaling) random variable.
Its expectation value is given by Eq.~\eqref{sortedInt}.
In that integral, let us introduce the \emph{next-neighbour distances}
$y_i = x_{i+1}-x_i$ ($i = 1,\ldots,n-1$) as variables; this results in
\begin{multline} \label{xyRawInt}
   \expect{F}  = n! \int_0^1 dx_1 \int_{0}^{1-x_1} dy_1
   \int_{0}^{1-x_1-y_1} dy_2
    \ldots \int_0^{1-x_1-\sum_{i=1}^{n-2} y_i} dy_{n-1}  \times 
    \\
     \times F(x_1,\, x_1+y_1,\, x_1+y_1+y_2,\, \ldots,\, x_1+\sum_{i=1}^{n-1}y_{i}).
\end{multline}
In the argument of $F$, we can certainly replace $x_1$ with $0$ due to 
the translation invariance of $F$. 
Moreover, we set
\begin{equation} \label{fHatDef}
  \hat F( y_1,\ldots,y_n ) = F(0,y_1,\ldots,\sum_{i=1}^{n-1} y_i),
\end{equation}
where the purpose of the apparently ``redundant'' variable $y_n$ 
is as follows: If we set $y_n = 1 - \sum_{i=1}^{n-1}y_i$, then
it is easily seen from the symmetry and translation invariance of $F$ 
that $\hat F$ is \emph{shift-symmetric} in 
the $n$ variables, in the sense that
\begin{equation} \label{shiftsymm}
  F(y_1,\ldots,y_n) = F(y_2,\ldots,y_n,y_1).
\end{equation}
Regarding the integration domain in Eq.~\eqref{xyRawInt}, we can see that 
the combined integration over $x_1,y_1,\ldots,y_{n-1}$ runs over the
$n$-dimensional standard simplex $V_n$; thus
\begin{equation}
   \expect{F}  = n! 
   \int_{V_{n}} dx_1\, d^{n-1}y \;
      \hat F(\yv).
\end{equation}
(The standard simplex and its properties are discussed in 
Appendix~\ref{SimplexApp}, which we will frequently refer to.)
Choosing a different coordinatization of the simplex, we can express
this as
\begin{equation}
   \expect{F}  = n! 
   \int_{V_{n-1}}  d^{n-1}y \;
   \int_0^{1-\sum_{i=1}^{n-1}y_i} dx_1 \;
      \hat F(\yv).
\end{equation}
The integration over $x_1$ can easily be executed:
\begin{equation}
   \expect{F}  = n! 
   \int_{V_{n-1}}  d^{n-1}y \;
   (1-\sum_{i=1}^{n-1} y_i)
      \hat F(\yv).
\end{equation}
Setting $y_n = 1- \sum_{i=1}^{n-1} y_i$, and comparing
with Eqs.~\eqref{surfaceIntNoDelta} and \eqref{surfaceIntDelta}
in Appendix \ref{SimplexApp},
we can rewrite this as an integral over the \emph{top surface}
$T_n$ of the simplex in $n$ dimensions:
\begin{equation}
   \expect{F}  = n! 
   \intcube  d^{n}y \, \delta(1-\sum_{i=1}^n y_i) \;
     y_n   \hat F(\yv).
\end{equation}
Now noting that the integration measure is completely symmetric
with respect to an exchange of variables, and using the shift-symmetry
of $F$ [cf. Eq.~\eqref{shiftsymm}], it is clear that we can replace
the factor $y_n$ in the integrand with any other $y_i$ without 
changing the integral's value; so we can as well replace it with the mean:
\begin{equation}
   \expect{F}  = n! 
   \intcube  d^{n}y \, \delta(1-\sum_{i=1}^n y_i) \;
     \frac{1}{n}\big(\sum_{i=1}^{n}y_i \big)   \hat F(\yv).
\end{equation}
However, under the integral, we have $\sum_{i=1}^{n}y_i=1$. Thus,
our result is
\begin{equation}
   \expect{F}  = (n-1)! 
   \intcube  d^{n}y \, \delta(1-\sum_{i=1}^n y_i) \;
       \hat F(\yv).
\end{equation}
Comparing with Proposition~\ref{tnEqualProp}, we can rewrite this as
\begin{equation} \label{tnEqualDist}
   \expect{F}  = 
   \intcube  d\muEqualT{n}(\yv)  \;
       \hat F(\yv).
\end{equation}
the next-neighbour coordinates are distributed equally (not independently!)
over the top surface $T_n$ of the standard simplex. Let us summarize:

\begin{thm}
Let $F$ be a symmetric and translation-invariant random variable
on $\Omega_n = [0,1]^n$, considered with the equal distribution.
Let $\hat F$ be the corresponding random variable  [see Eq.~(\ref{fHatDef})]
on $\Omega_n' = T_n$, considered with the equal distribution on $T_n$.
Then
\begin{equation*}
	\expect{F} = \expect{\hat F}.
\end{equation*}
\end{thm}

In fact,
it will be more convenient in most cases to define the 
random variables directly in terms of the next-neighbour
coordinates; given that the so-defined variable $\hat F$ is shift-symmetric,
we can always define an underlying symmetric and translation-invariant
random variable $F$. We will not even distinguish 
the two associated random variables in notation (where this is unambiguous).

\subsection{Connectedness}

We will now turn to calculate the probability that the MANET
is connected. This needs some explanation with regard to the
periodic boundary conditions: We will call the MANET connected if 
\emph{all} next neighbours are connected, including the
left-most and the right-most one (which are connected 
``via the boundary''). More formally, we define the 
event $\connpb$ in next-neighbour coordinates as
\begin{equation} \label{connpbDef}
   M_\connpb := \{ \yv \in T_n \, | \, \forall j: \, y_j < \rho \}.
\end{equation} 
We also consider the more general event $\discpb{k}$ for $k \in \nbb_0$, defined as
\begin{equation} \label{discpbDef}
   M_\discpb{k} := \{ \yv \in T_n \, | \, y_j \geq \rho  \text{ for exactly $k$ values of $j$} \},
\end{equation} 
meaning that the network is disconnected at $k$ places (or, equivalently
speaking, into $k$ segments). Note that $\connpb=\discpb{0}$.

Our task is to calculate the probability of $\discpb{k}$. 
A central tool for this is the inclusion-exclusion formula
(see Appendix~\ref{IncExcApp}); it gives us
\begin{equation} \label{discIncExc}
  P_\discpb{k}= \sum_{j=k}^{n} (-1)^{j-k} \binom{j}{k}  S_j,
\end{equation}
where
\begin{equation} \label{discIncExcSum}
   S_j = \sum_{ \{m_1,\ldots,m_j\} } P( y_{m_1} \geq \rho \wedge \ldots 
         \wedge y_{m_j} \geq \rho )  .
\end{equation}
It remains to calculate the probability of the event 
under the sum, which is handled in the following lemma.

\begin{lemm} \label{pGeqLemm}
Let $j \in \{0,\ldots,n\}$, and let 
$\{m_1,\ldots,m_j\} \subset \{ 1, \ldots, n \}$ be
a $j$-element subset. Then
\begin{equation*}
P( y_{m_1} \geq \rho \wedge \ldots 
         \wedge y_{m_j} \geq \rho  )
 = \begin{cases}
 	(1- j \rho)^{n-1} & \text{if }j \leq 1/\rho, 
 	\\
 	0	& \text{otherwise}. 
 \end{cases}   
\end{equation*}

\end{lemm}

\begin{proof}
Let $\hat P$ be the probability in question. 
We will prove the result by induction on $j$.
For $j=0$, it is obvious that $\hat P=1$ as proposed. 
So assume that we have
verified the result for $j-1$ in place of $j$. 
The case $j > 1/\rho$ is obvious, since $\sum_{i=1}^{n} y_i = 1$;
so let $j \leq 1/\rho$ in the following. The characteristic function of the event
can be expressed as a product of $\theta$ functions;\footnote{
See Eq.~\eqref{thetaDef} in Appendix~\ref{SimplexApp} for the definition of the Heaviside $\theta$ function.}
that results in
\begin{equation} \label{pxInt}
  \hat P = \intcube  d\muEqualT{n}(\yv) \;
  \prod_{i=1}^{j} \theta(y_{m_i}-\rho).
\end{equation}
Applying Lemma~\ref{tnScalingLemm} with respect to the variable $y_{m_j}$,
we obtain
\begin{multline}
  \hat P 
  = (1-\rho)^{n-1} \intcube  d\muEqualT{n}(\yv) \;
  \prod_{i=1}^{j-1} \theta(y_{m_i}-\frac{\rho}{1-\rho})
  \\
  = (1-\rho)^{n-1} P( y_{m_1} \geq\frac{\rho}{1-\rho} \wedge \ldots 
         \wedge y_{m_{j-1}} \geq \frac{\rho}{1-\rho}  ).
\end{multline}
Here we can apply the induction hypothesis 
for $j-1$ in place of $j$ and $\rho/(1-\rho)$ in place of $\rho$; note that the condition $j \leq 1/\rho$
guarantees that $j-1 \leq (1-\rho)/\rho$.
This shows us that
\begin{equation}
  \hat P =  (1-\rho)^{n-1} \; \big(1 - (j-1)\frac{\rho}{1-\rho} \big)^{n-1} = (1-j\rho)^{n-1},
\end{equation}
which proves the lemma.
\end{proof}

Applying this lemma in Eq.~\eqref{discIncExcSum}, and then inserting into
Eq.~\eqref{discIncExc},
we can establish an explicit expression for $P_\discpb{k}$. Note that in
Eq.~\eqref{discIncExcSum}, all summands are in fact equal, so that
we only need to count the number of terms, which is $\binom{n}{j}$.
Our result then is:

\begin{thm} \label{pdiscPeriodicThm}
In the 1-dimensional MANET with periodic boundary conditions,
one has for each $k \in \nbb_0$,
\begin{equation*}
  P_\discpb{k} = \sum_{j=k}^{[1/\rho]} (-1)^{j-k} \binom{j}{k} \binom{n}{j} (1-j\rho)^{n-1}.
\end{equation*}
In particular,
\begin{equation*}
  P_\connpb = \sum_{j=0}^{[1/\rho]} (-1)^{j} \binom{n}{j} (1-j\rho)^{n-1}.
\end{equation*}
\end{thm}

Here $[1/\rho]$ is the Gauss bracket of $1/\rho$, i.e. the greatest integer
which is less or equal to $1/\rho$. Note that the formula is valid
for $[1/\rho]>n$ as well, since the factor $\binom{n}{j}$ evaluates to $0$
for $j>n$, so that these summands automatically vanish.

We have thus found an explicit expression for $P_\discpb{k}$;
the expression is defined piecewise as a polynomial in $\rho$
of degree $n-1$. In particular for small values of $\rho$,
the sum involves terms of high modulus and opposite sign;
thus a numerical evaluation with floating-point techniques
may lead to problems due to round-off errors. However, 
inserting $\rho$ as a fraction, we can use integer arithmetics
in order to evaluate the sum, thus bypassing the problems mentioned.

\begin{figure} 
\centering
    \resizebox{0.7\textwidth}{!}{
      \includegraphics{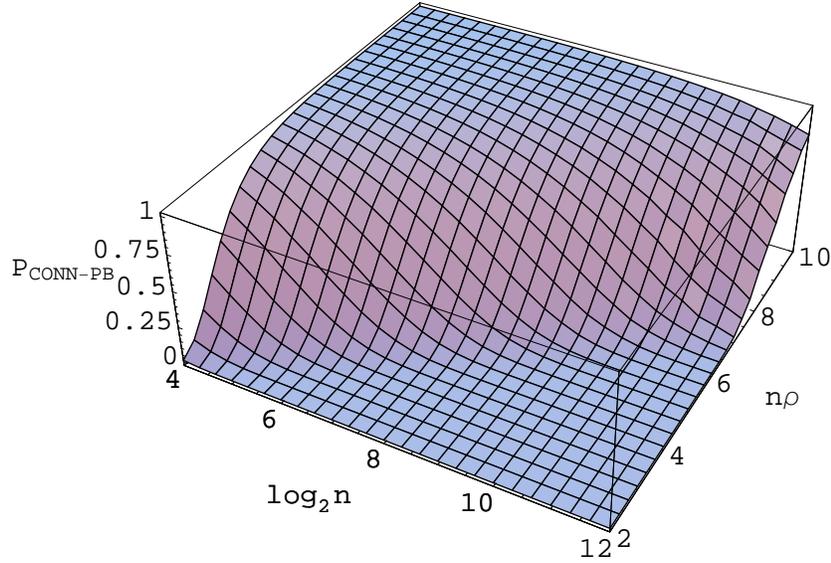}
    }  
\caption{Probability of connectedness for the 1-dimensional MANET} \label{PConnFig}
\end{figure}

Figure~\ref{PConnFig} shows the behaviour of $P_\connpb$,
plotted against $n$ (on a logarithmic scale) and $n \rho$.
Two things are noticeable: First, at fixed $n$, we obviously
have $P_\connpb \to 1$ for $\rho \to \infty$ and $P_\connpb \to 1$ for $\rho \to 0$.
This is expected and can directly be seen from the arithmetic
expressions. Second, it seems that in the limit or large $n$,
the probability $P_\connpb$ is basically a function of
one parameter $n\rho - \ln n$. This asymptotic behaviour
will be discussed in the next section.

\subsection{Asymptotic behaviour} \label{connAsymptoticSec}

Apart from the probability of connectedness for fixed parameters
$\rho$ and $n$, we are particularly interested in the behaviour
of our model for large MANETs, that is, in the limit $n \to \infty$.
However, although Fig.~\ref{PConnFig} suggests that there is 
some well defined large-scale limit of the system, it is not apparent
from Theorem~\ref{pdiscPeriodicThm} how $P_\discpb{k}$ behaves in this limit.
In this section, we will discuss $P_\discpb{k}$ in the large-scale
limit and derive an asymptotic approximation formula.

Since the detailed calculation turns out to be quite technical,
let us first present a heuristic sketch of the underlying ideas,
where we will restrict ourselves to $P_\connpb$.
We can rewrite the expression from Theorem~\ref{pdiscPeriodicThm}
as
\begin{equation} \label{pconnHeuristicSum}
  P_\connpb = \sum_{j=0}^{[1/\rho]} (-1)^{j} \frac{n!}{j! (n-j)!} (1-j\rho)^{n-1}.
\end{equation}
Using Sterling's formula ($\ln n! \approx n \ln n$) and Taylor expansion
($\ln (1-x) \approx -x$), we see that for large $n$ and moderate $j$,
\begin{equation}
	\ln  \frac{n!}{j! (n-j)!}  \approx j (\ln n - \ln j),
	\quad
	\ln (1-j\rho)^{n-1} \approx -j \rho n;
\end{equation}
so the polynomial factor $(1-j\rho)^{n-1}$ dominates the binomial factor
for medium to large $j$, such that only a very limited number of summands
$(j \leq j_0)$ will actually contribute to the sum 
in Eq.~\eqref{pconnHeuristicSum}. For these terms, we can individually
let $n \to \infty$ at fixed $j$. Here we have
\begin{equation}
	\frac{n!}{(n-j)!} = n (n-1) \ldots (n-j+1) \approx n^j
	\quad \text{and} \quad
	(1-j\rho)^{n-1} \approx e^{-j\rho n}. 	
\end{equation}
Inserting into Eq.~\eqref{pconnHeuristicSum}, this means that
\begin{equation} \label{pconnHeuristicApprox}
  P_\connpb \approx \sum_{j=0}^{j_0} \frac{(-1)^{j}}{j!} (n e^{-n\rho})^j.
\end{equation}
Without changing the value of the sum significantly, we can replace $j_0$
with $\infty$ here; then the sum becomes an exponential series, and we see that
\begin{equation} \label{pconnHeuristicLimit}
  P_\connpb \approx \exp( n e^{-n\rho}) = \exp(-\exp(-n\rho+\ln n)).
\end{equation}

Of course, controlling the limit $n \to \infty$ is in fact not as easy as suggested above,
and we have to turn the heuristic arguments into a rigorous proof
in order to be sure about the large-scale behaviour of the model.
This is the content of the following theorem.

\begin{thm} \label{pconnPbAsympThm}
Let $(\rho_n)$ be a sequence in $\rbb^+$, and suppose there is an $\eta \in \rbb$
such that
\begin{equation*}
	n \rho_n - \ln n \longnlim \eta.
\end{equation*}
Then, we have for every $k \in \nbb_0$:
\begin{equation*}
	P_\discpb{k}^{(n,\rho_n)} \longnlim \frac{e^{-\eta k}}{k!} \exp(-e^{-\eta}).
\end{equation*}
\end{thm}

\begin{proof}
First, let us note some properties of the specified limit:
Since $n \rho_n - \ln n \to \eta$, we certainly have 
$\rho_n \sim \ln n / n$ and thus
\begin{equation}  \label{qnlim}
  \rho_n \to 0, \quad
  n  \rho_n \to \infty, \quad
  n \rho_n^2 \to 0.
\end{equation}
Now let us turn to $P_\discpb{k}$. We can rewrite the expression from
Theorem~\ref{pdiscPeriodicThm} as
\begin{equation}
  P_\discpb{k} = \frac{n^k e^{-k n \rho_n}}{k!} 
  \sum_{j=k}^{[1/\rho]} (-1)^{j-k} \frac{ n! \, n^{-k} }{ (n-j)!\, (j-k)! } 
   (1-j\rho_n)^{n-1} e^{kn\rho_n}.
\end{equation}
Shifting the summation index by $-k$, this is equivalent to
\begin{equation} \label{sumWithADef}
  P_\discpb{k} = \frac{n^k e^{-k n \rho_n}}{k!} 
  \sum_{j=0}^{[1/\rho]-k} \frac{(-1)^{j}}{j!} 
   \underbrace{ \frac{ n! \, n^{-k} }{ (n-j-k)! } 
    (1-(j+k)\rho_n)^{n-1} e^{kn\rho_n} }_{=: a_j}.
\end{equation}
In the factor that precedes the sum, it is obvious that
\begin{equation}
    n^k e^{-k n \rho_n} = e^{-k (n\rho_n - \ln n)} \longnlim e^{-\eta k};
\end{equation}
so it remains only to control the convergence of the sum itself.
We will next investigate how fast the summand terms $a_j$ 
vanish for large $j$. We can certainly say that
\begin{equation}
  \frac{n!}{(n-j-k)!} = n (n-1) \cdots (n-j-k+1) \leq n^{j+k}
\end{equation}
and thus
\begin{equation}
  \ln a_j \leq j \ln n 
  + (n-1) \ln(1-(j+k)\rho_n) + k n \rho_n.
\end{equation}
Since it is known from the Taylor series 
of $\ln(1-x)$ that  
$\ln(1-x) \leq -x$ for all $x \in (-\infty,1)$, we see that
\begin{equation}
  \ln a_j \leq j (\ln n  - n \rho_n + \rho_n) + k \rho_n.
\end{equation}
Now since $n \rho_n - \ln n \to \eta$ and $\rho_n \to 0$, we can certainly find
$n_0$ such that 
\begin{equation} \label{ajbounds}
  \forall n \geq n_0 \; \forall j: \;
  \ln a_j \leq 2 \eta j + 1 \quad
  \text{or, equivalently,}
  \quad
  a_j \leq e^{2 \eta j +1}.
\end{equation}
According to Stirling's formula (cf. Theorem~\ref{stirlingThm} in Appendix~\ref{LimitApp}), 
we can say that for any $j$
\begin{equation}
  j! \geq \big(  \frac{j}{e} \big)^j,
\end{equation}
and thus for $n \geq n_0$
\begin{equation}
  \frac{a_j}{j!} \leq e  \cdot \big(  \frac{e^{2 \eta + 1}}{j} \big)^j .
\end{equation}
Given $\epsilon > 0$, we can thus find $j_0$ such that
\begin{equation}
  \forall n \geq n_0 \; \forall j \geq j_0: \;
  \frac{a_j}{j!} \leq   \frac{\epsilon}{2^j}.
\end{equation}
This means that
\begin{equation} \label{ajRemainder}
  \big| \sum_{j=j_0}^{ [1/\rho_n]-k} \frac{(-1)^j}{j!} a_j  \big|
  \leq \epsilon \sum_{j=j_0}^{ \infty } \frac{1}{2^j} \leq 2 \epsilon.
\end{equation}
Moreover, after possibly increasing $j_0$, we can achieve that
\begin{equation} \label{expRemainder}
  \big| \sum_{j=j_0}^{ \infty } \frac{(-1)^j}{j!} e^{-\eta j}  \big|
  \leq \epsilon ,
\end{equation}
since the exponential series $\sum_j x^j/j!$ converges absolutely on $\rbb$.
Further, we can assume that $1/\rho_n > j_0+k$ for $n \geq n_0$.

It remains to estimate the convergence of the terms for $j < j_0$
in Eq.~\eqref{sumWithADef}. To that end, note that the above estimates
are uniform in $n$: Once we have fixed
$j_0$ and $n_0$ for given $\epsilon$, we can consider the limit
$n \to \infty$ without changing $j_0$. Thus, there are only finitely
many terms left to estimate, and we can consider the limit in 
each of them individually: We want to show that for each $j < j_0$,
one has
\begin{equation} \label{ajlim}
	a_j / e^{- \eta j} \longnlim 1.
\end{equation}
Explicitly, we know that
\begin{equation} 
	a_j / e^{- \eta j} = \frac{n! \; n^{-k-j}}{(n-j-k)!} \; 
	n^j \;
	    (1-(j+k)\rho_n)^{n-1} e^{kn\rho_n + \eta j}.
\end{equation}
Certainly, the first factor converges as $n \to \infty$:
\begin{equation} 
\frac{n! \;n^{-k-j}}{(n-j-k)!} =
\frac{n (n-1) \cdots (n-j-k+1)}{n^{j+k}} \longnlim 1.
\end{equation}
Furthermore, we see that
\begin{equation} \label{ajpart}
\ln \big(
n^j \;
	    (1-(j+k)\rho_n)^{n-1} e^{kn\rho_n + \eta j}
\big)
= j \ln n + (n-1) \ln (1 - (j+k) \rho_n) + kn\rho_n + \eta j.
\end{equation}
Again, we use the Taylor expansion $\ln (1-x) = -x + O(x^2)$;
this results in
\begin{equation} \label{secondFactorConv}
%\ln \big(
%n^j \;
%	    (1-(j+k)\rho_n)^{n-1} e^{kn\rho_n + \eta j}
%\big)
\eqref{ajpart}
= j (\ln n - n \rho_n + \eta) + O(\rho_n) + O(n \, \rho_n^2).
\end{equation}
According to Eq.~\eqref{qnlim}, all of the terms on the right-hand side
vanish in the limit; this proves Eq.~\eqref{ajlim}.
Since we had seen in Eq.~\eqref{ajbounds} that the $a_j$ are
uniformly bounded in $n$ (at fixed $j$), we have \emph{a forteriori}
that
\begin{equation} \label{ajexpEstimate}
  | a_j - e^{-\eta j} | \leq\, |a_j|\, | 1 - \frac{e^{-\eta j}}{a_j} | \, \longnlim 0.
\end{equation}
This means that we can find $n_1 \geq n_0$ such that for
any $n \geq n_1$,
\begin{equation} \label{ajexpSumEstimate}
 \big| \sum_{j=0}^{j_0-1} \frac{(-1)^j}{j!} a_j -
   \sum_{j=0}^{j_0-1} \frac{(-1)^j}{j!} e^{-\eta j} \big| 
   \leq \epsilon.
\end{equation}
Now combining Eqs.~\eqref{ajRemainder}, \eqref{expRemainder}, 
and \eqref{ajexpSumEstimate}, we know that
\begin{equation}
 \forall \epsilon > 0 \;
 \exists n_1 \;
 \forall n \geq n_1:
 \big| \sum_{j=0}^{[1/\rho_n]-k} \frac{(-1)^j}{j!} a_j -
   \sum_{j=0}^{\infty} \frac{(-1)^j}{j!} (e^{-\eta})^{j} \big| 
      \leq 5 \epsilon.
\end{equation}
Rewriting the exponential series as an exponential function, this means
\begin{equation}
  \sum_{j=0}^{[1/\rho_n]-k} \frac{(-1)^j}{j!} a_j 
  \longnlim
   \exp(- e^{-\eta}) .
\end{equation}
Inserted into Eq.~\eqref{sumWithADef}, this proves the theorem.
\end{proof}

Let us add another result for the limit $n \to \infty$,
which we state for $P_\connpb$ only: Suppose
that $n\rho_n - \ln n \to \infty$ in the limit. Then for
given $\eta$, we can certainly construct a sequence
$(\rho_n')$ with $\rho_n' < \rho_n$ such that $n \rho_n' - \ln n \to \eta$.
Since $P_\connpb^{(n,\rho)}$ is monotonous in $\rho$ at fixed $n$,
we see that
\begin{equation}
  P_\connpb^{(n,\rho_n)} \geq  P_\connpb^{(n,\rho_n')} \longnlim \exp(-e^{-\eta}).
\end{equation}
We can choose $\eta$ arbitrarily high here; that means
$P_\connpb^{(n,\rho_n)} \to 1$. A similar result for 
$n\rho_n - \ln n \to -\infty$ can be obtained in the same way.
Let us note this for reference:

\begin{thm} \label{pconnPbExtThm}
Let $(\rho_n)$ be a sequence in $\rbb^+$, and suppose that
\begin{equation*}
	n \rho_n - \ln n \longnlim + \infty
	\quad \text{or} \quad 
	n \rho_n - \ln n \longnlim - \infty.
\end{equation*}
Then, we have 
\begin{equation*}
	P_\connpb^{(n,\rho_n)} \longnlim 1
	\quad \text{or, respectively,}  \quad
	P_\connpb^{(n,\rho_n)} \longnlim 0.
\end{equation*}
\end{thm}

(A similar result could be proved for $P_\discpb{k}$, but we will make
no use of it.)

\begin{figure} 
\centering
\mbox{
  \subfigure[absolute]  {
    \resizebox{0.45\textwidth}{!}{
      \includegraphics{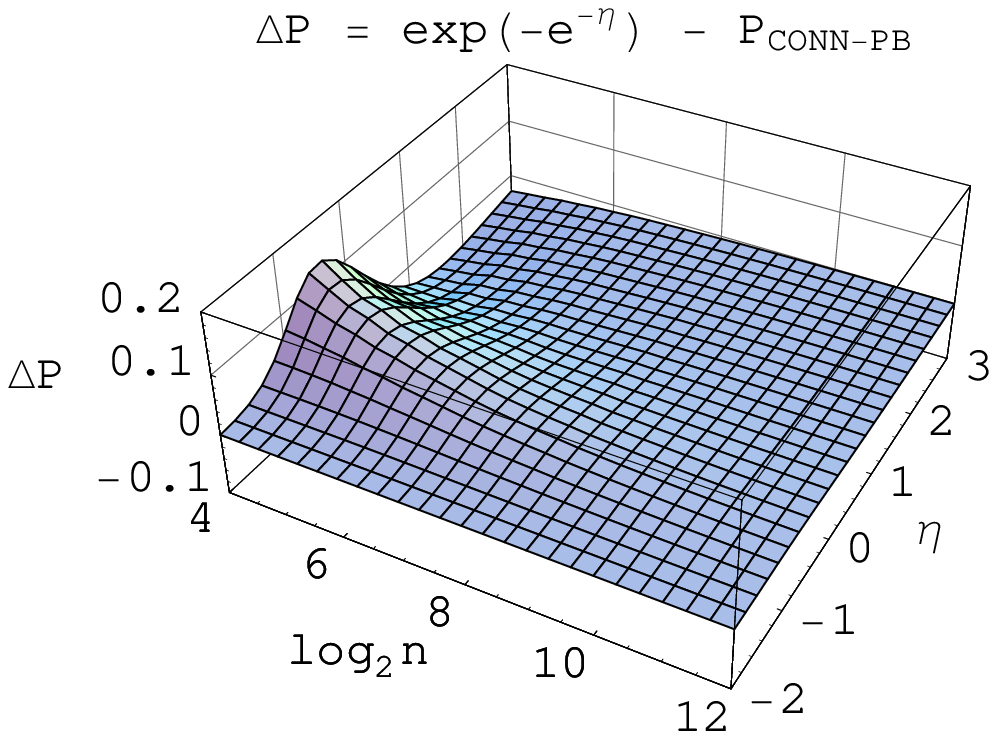}
    }
  }
  \subfigure[relative]  {
    \resizebox{0.45\textwidth}{!}{
      \includegraphics{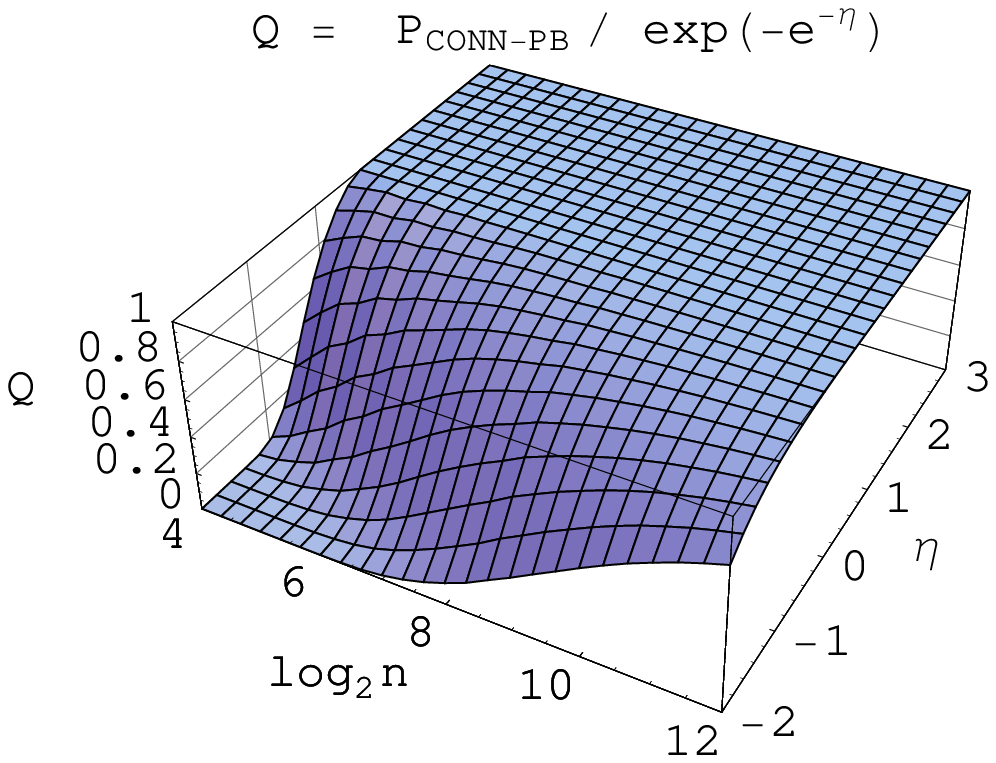}
    }  
  }
}
\caption[Quality of the asymptotic approximation for $P_\connpb$]
{Quality of the asymptotic approximation for $P_\connpb$. 
Relative and absolute comparisons are plotted against $n$ (on a logarithmic scale)
and $\eta = n \rho - \ln n$.}
\label{approxFig}
\end{figure}

Figure~\ref{approxFig} shows the quality of the asymptotic approximation
of $P_\conndb$ for growing $n$. As might be expected from the details of the proof,
the convergence is particularly fast for large $\eta$. For $\eta<0$ and low $n$, 
the absolute error is small, but on a relative scale, the approximation 
is rather unusable. This may be understood from the fact that $P_\conndb$
is exactly $0$ for $\rho < 1/n$, as is apparent from the model, while
the approximation $\exp(-e^{-\eta})$ still gives positive, if very small, values.

\section{Connectivity with disconnected boundary conditions} \label{disconnectedSec}

We will now aim at transferring our results to the case of
``disconnected boundaries,'' i.e. where no connections
between the left-most and the right-most node are possible.
This is the situation considered by Santi and Blough \cite{SanBlo:transmitting_range},
and one of our goals is to compare our results to theirs.
We wish to show that the specific form of
boundary conditions has no effect in the limit $n \to \infty$,
that is, our results from Sec.~\ref{connAsymptoticSec}
hold for disconnected boundary conditions, too.

\subsection{Estimates} \label{disconnEstimatesSec}

Both models -- the MANET with periodic and disconnected boundary conditions --
are formulated on the same probability space, but are based on different
events and random variables. For the disconnected boundaries, we 
consider the events $\discdb{k}$, defined as
\begin{equation} \label{discdbDef}
  M_\discdb{k} = \{  \xv \in [0,1]^n  \,|\, x_{i+1}' - x_{i}' \geq \rho \text{ for exactly $k$ values of }
                     i \in \{1,\ldots,n-1\} \}
\end{equation}
where $x'_1 \leq x'_2 \leq \ldots$ are the sorted  coordinates $(x_i)$;
as discussed, these events differ from 
$\discpb{k}$ only by the handling of the nodes on the boundary.
The event $\discdb{k}$, or rather its characteristic function,
is certainly scaling and also symmetric, since it only refers to the 
sorted coordinates. However, it is no longer translation invariant.
Thus we can apply Eq.~\eqref{sortedInt}, but no longer the results of
Sec.~\ref{transformSec}. For example, we might calculate the probability
of connectedness, i.e. of the event $\conndb=\discdb{0}$, as
\begin{equation} \label{sortedIntConn}
   P_\conndb = n! \int_0^1 dx_1 \int_{x_1}^{\min(1,x_1+\rho)} dx_2 
   \ldots \int_{x_{n-1}}^{\min(1,x_{n-1}+\rho)} dx_n.
\end{equation}
This integral can in principle be solved (for fixed $n$) and gives a 
piecewise-defined polynomial in $\rho$ of degree $n$. (One might e.g.
use computer algebra to solve it for realistic $n$.) However, a 
closed solution for arbitrary $n$ seems to be out of reach.
Instead, we will restrict to estimates of the difference between
the periodic and disconnected boundary conditions.

It is obvious that $M_\discpb{k} \subset M_\discdb{k}$; however, the opposite
inclusion is not true: A configuration $\xv \in M_\discdb{k}$ might be
disconnected at $k+1$ places with respect to periodic boundaries, 
namely if the left-most point in $\xv$ is not connected to 
the right-most one ``via the boundary''. More precisely, let
\begin{equation}
  S := \{ \xv \in [0,1]^n \,|\, x_1' + 1 - x_n' \geq \rho ,
  \text{ where } x_1' = \min\{x_i\}, \; x_n' = \max\{x_i\} \};
\end{equation}
this is the event that the network is disconnected ``at the boundary''.
Then it is clear that
\begin{equation}
M_\discdb{k} = M_\discpb{k} \cup (M_\discpb{(k+1)} \cap S),
\end{equation}
where the union is disjoint. This gives us the following inequality:
\begin{equation}
 P_\discpb{k} \leq P_\discdb{k} \leq P_\discpb{k} + P(M_\discpb{(k+1)} \cap S).
\end{equation}
As an estimate, we can certainly say that
\begin{equation} \label{allKEstimate}
 P_\discpb{k} \leq P_\discdb{k} \leq P_\discpb{k} + P(S);
\end{equation}
it then only remains to calculate $P(S)$. 

The event $S$ is most conveniently described in sorted coordinates;
in fact, if $x_1 \leq \ldots \leq x_n$, we can write its characteristic function as
\begin{equation}
      \chi_S (x_1,\ldots,x_n) = \theta(x_1+(1-\rho)-x_n).
\end{equation}
Using Eq.~\eqref{sortedInt} to calculate the expectation value, we get
\begin{equation} \label{sortedIntS}
   P_S = n! \int_0^1 dx_1 \int_{x_1}^1 dx_2 \ldots \int_{x_{n-1}}^1 dx_n \;
   \theta(x_1+(1-\rho)-x_n).
\end{equation}
Let us split the integral in a sum $P_S = I_1 + I_2$, where $I_1$
covers the integration domain $(\rho,1)$ in the variable $x_1$,
and $I_2$ covers the interval over $(0,\rho)$. If $x_1 \in (\rho,1)$,
then the argument of the theta function is always positive, thus
\begin{equation} 
   I_1 = n! \int_\rho^1 dx_1 \int_{x_1}^1 dx_2 \ldots \int_{x_{n-1}}^1 dx_n \;1.
\end{equation}
Introducing new variables $z_1 = (x_1-\rho)/(1-\rho)$, and $z_i = (1-x_i)/(1-\rho)$
for $i = 2,\ldots,n$, this reads 
\begin{equation} 
   I_1 = n! \;(1-\rho)^n \int_0^1 dz_1 
   \int_{0}^{1-z_1} dz_2 
   \ldots 
   \int_0^{1-\sum_{i=1}^{n-1} z_i} dz_n \,
   = 
   n! \; (1-\rho)^n \vol(V_n). 
\end{equation}
The volume of the $n$-dimensional standard simplex is known from
Eq.~\eqref{volvn}; our result thus is $I_1 = (1-\rho)^n$. Now for the second
integral, namely
\begin{equation} 
   I_2 = n! \int_0^\rho dx_1 
   \int_{x_1}^1 dx_2 \ldots \int_{x_{n-1}}^1 dx_n \;
   \theta(x_1+(1-\rho)-x_n).
\end{equation}
Here we introduce new variables $z_i = (x_{i+1}-x_i)/(1-\rho)$
for $i = 1,\ldots,n-1$; this leads us to
\begin{equation} 
   I_2 = n! \;(1-\rho)^{n-1} 
   \int_0^\rho dx_1 
   \int_0^{ \frac{1-x_1}{1-\rho}  } dz_1 
   \int_0^{ \frac{1-x_1}{1-\rho} - z_1 } dz_2
   \ldots 
   \int_0^{ \frac{1-x_1}{1-\rho} - \sum_{i=1}^{n-2} z_i } dz_{n-1} \;
   \theta(1- \sum_{i=1}^{n-1} z_i).
\end{equation}
Note that the $\theta$ function restricts the domain of integration for $\zv$
to the $(n-1)$-dimensional standard simplex, which is 
completely covered by the integration since $(1-x_1)/(1-\rho) > 1$.
Thus
\begin{equation}
   I_2 = n! \;(1-\rho)^{n-1} 
   \; \rho \; \vol(V_{n-1}) 
   = n \; \rho \;(1-\rho)^{n-1}.
\end{equation}
Combining the results for $I_1$ and $I_2$ in Eq.~\eqref{allKEstimate},
our result is:

\begin{lemm} \label{dbEstimateLemm}
Let $k \in \nbb_0$. Then
\begin{equation*}
 P_\discpb{k} \leq P_\discdb{k} \leq P_\discpb{k} + (1-\rho)^n + n\rho\;(1-\rho)^{n-1}.
\end{equation*}
\end{lemm}

This estimate is certainly not very strict
and might be improved, but it is already sufficient for our purposes:
Note that in the limit $n \to \infty$ and $\rho \to 0$, we have
$\ln((1-\rho)^n) = -n\rho + O(n\rho^2)$, and $\ln (n\rho(1-\rho)^{n-1}) = 
\ln ( n \rho ) - n \rho + O(\rho) + O(n\rho^2)$; thus we see from Eq.~\eqref{qnlim}
that the difference between upper bounds and lower bounds
in Lemma~\ref{dbEstimateLemm} vanishes as $n\rho - \ln n \to \eta$.
This means that we can directly transfer the results from 
Theorem~\ref{pconnPbAsympThm} to the case of disconnected boundary conditions. 
It is also straightforward to transfer the results for 
$n\rho - \ln n \to \pm \infty$ from Theorem~\ref{pconnPbExtThm}. 
Let us summarize this as a separate statement.

\begin{thm} \label{pconnDbAsympThm}
Let $(\rho_n)$ be a sequence in $\rbb^+$, and suppose there is an $\eta \in \rbb$
such that
\begin{equation*}
	n \rho_n - \ln n \longnlim \eta.
\end{equation*}
Then, we have for every $k \in \nbb_0$:
\begin{equation*}
	P_\discdb{k}^{(n,\rho_n)} \longnlim \frac{e^{-\eta k}}{k!} \exp(-e^{-\eta}).
\end{equation*}
In particular,
\begin{equation*}
	P_\conndb^{(n,\rho_n)} \longnlim \exp(-e^{-\eta}).
\end{equation*}
In the case 
\begin{equation*}
	n \rho_n - \ln n \longnlim + \infty
	\quad \text{or} \quad 
	n \rho_n - \ln n \longnlim - \infty,
\end{equation*}
one has
\begin{equation*}
	P_\conndb^{(n,\rho_n)} \longnlim 1
	\quad \text{or, respectively,}  \quad
	P_\conndb^{(n,\rho_n)} \longnlim 0.
\end{equation*}
\end{thm}

Overall, this makes our claim precise that the choice of 
boundary conditions does not play a role in the limit
$n \to \infty$.

\subsection{Comparison with the literature} \label{pconnLiteratureSec}

Now that we have established our results for the system with
disconnected boundary conditions, we are in the position 
to compare them with 
existing results in the literature -- in particular with
those of Santi et~al. 
\cite{SBV:range_assignment,SanBlo:connectivity,SanBlo:transmitting_range}
who investigated the probability of connectedness
using the same mathematical model, obtaining
analytical estimates (with different techniques than ours) 
and also numerical results. 

We start with the analytical results. 
For comparison purposes, let us first state the following
special case of Theorem~\ref{pconnDbAsympThm}. (We return
to the parameters $r$ and $\ell$ in place of $\rho = r/\ell$ here.)
\begin{corr} \label{SantiCompareCorr}
	Consider the 1-dimensional MANET model with 
	parameters $n$, $r=r(n)$, $\ell=\ell(n)$ and disconnected 
	boundary conditions.
	If there is an $\epsilon >0$ such that for large $n$,
	\begin{equation*}
		nr \geq (1+\epsilon)\; \ell \ln n,
		\quad \text{then} \quad
		P_\conndb \longnlim 1.
	\end{equation*}
	If, on the other hand, 
	\begin{equation*}
		nr \leq (1-\epsilon)\; \ell \ln n,
		\quad \text{then} \quad
		P_\conndb \longnlim 0.
	\end{equation*}
\end{corr}
For the same situation, Santi and Blough \cite[Theorem~7]{SanBlo:transmitting_range} 
state that, when expressed in our notation,
\begin{enumerate}
  \renewcommand{\theenumi}{(\alph{enumi})}
  \renewcommand{\labelenumi}{\theenumi}

  \item \label{SantiK}
   		if $ n r = k \;\ell \ln \ell$ with some $k >2$, then $P_\conndb \to 1$,
  \item if $ n r = 2 \;\ell \ln \ell$ and $r(n) \to \infty$, then also $P_\conndb \to 1$,
  \item if $ n r = k \;\ell \ln \ell$ with $k \leq (1-\epsilon)$ for some $0 < \epsilon < 1$ 
  		and $r \in \Theta(\ell^\epsilon)$, then $P_\conndb \not\to 1$,
  \item \label{SantiZero}
  		if $( n r ) / ( \ell \ln \ell ) \to 0$, then $P_\conndb \not\to 1$,
\end{enumerate}
in the limit $n \to \infty$ and $\ell \to \infty$,
under the additional condition that
$r/\ell \to 0$.
While these results are formally quite similar to ours, they are only compatible
with them when the factor $\ln \ell$ does not differ significantly from $\ln n$.
In practice, one will usually consider the case $r = const.$, and here
in fact no difference arises. However, from a mathematical standpoint, this 
is not implied by the conditions of the theorem; in fact, one may construct
cases where the predictions of Corollary~\ref{SantiCompareCorr} are in 
conflict with the results from \cite{SanBlo:transmitting_range}.
For example, consider the case $n = k \; \ell^{k+1}$, $r = \ell^{-k} \ln \ell$,
where $k > 2$.
Then $P_\conndb \to 1$ according to \ref{SantiK}, but 
$P_\conndb \to 0$ according to Corollary~\ref{SantiCompareCorr}.
On the other hand, let $\ell = e^n$, $r = e^n/ \ln n$.
Then $P_\conndb \to 1$ according to Corollary~\ref{SantiCompareCorr},
in contradiction to~\ref{SantiZero}.

The present author claims that these differences are due to the fact
that the arguments 
presented in \cite{SBV:range_assignment,SanBlo:connectivity,SanBlo:transmitting_range}
are inconclusive.
The reader will find a detailed discussion hereof in Appendix~\ref{SantiApp}.

Another analytical result for connectedness was obtained 
by Piret~\cite{Pir:radio_connectivity} in a similar situation: He modeled 
the nodes of a 1-dimensional MANET by a Poisson process of constant density
$d$, and proved that for the radio range set to 
\begin{equation} \label{rPiret}
   r = k \frac{\ln(\ell d)}{2d}
\end{equation}
with a constant $k>0$, one has
\begin{equation} \label{piretResult}
   P_\text{Connectedness} \xrightarrow[]{\ell \to \infty} 
    \begin{cases}
        1  \quad &  \text{if } k>2, \\
        0       &  \text{if } k<2.
    \end{cases}
\end{equation}
Due to the use of a Poisson process, the node numner $n$ is a 
random variable in this case and not a fixed number; however,
for large $\ell$ (and hence $n$) one would expect that $n$ assumes
its mean value $\ell d$ with low variance. In fact, setting 
$n = \ell d$ in Eq.~\eqref{rPiret}, the result \eqref{piretResult}
is just what Corollary~\ref{SantiCompareCorr} amounts to.

\begin{figure} 
\centering
\begin{minipage}{0.76\textwidth}
  \resizebox{\textwidth}{!}{
  \includegraphics{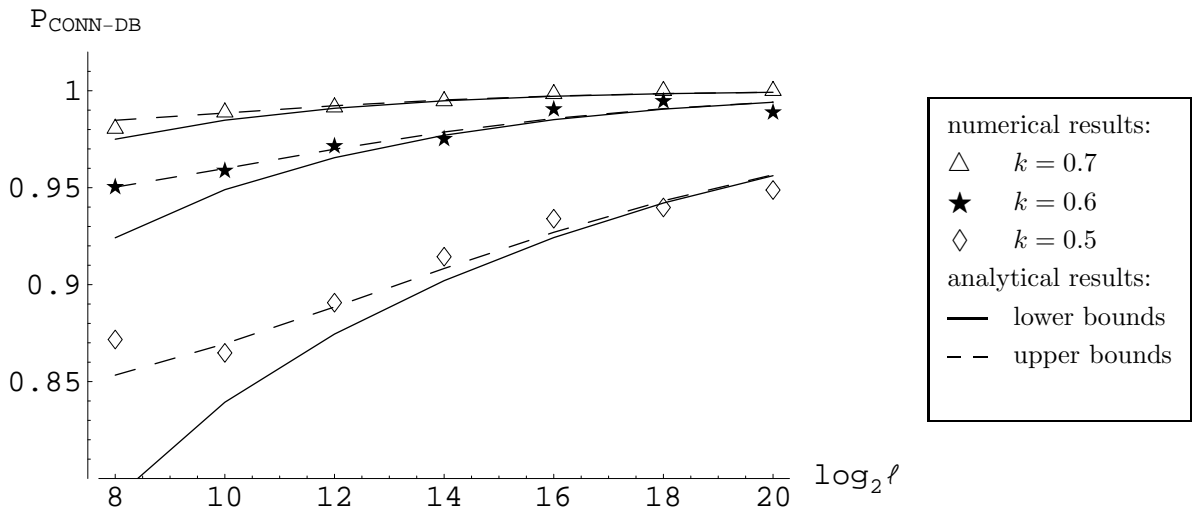}
  }
\end{minipage}
\fbox{
  \begin{minipage}{0.15\textwidth}
  \begin{tabbing}
  \footnotesize{numerical results:} \\
  \footnotesize{$\triangle$} \;\quad \= \footnotesize{$k=0.7$} \\
  \footnotesize{$\bigstar$}  \> \footnotesize{$k=0.6$} \\
  \hspace{1.5pt}\footnotesize{$\lozenge$}  \>  \footnotesize{$k=0.5$} \\
  \footnotesize{analytical results:} \\
  \rule[2pt]{15pt}{0.5pt} \>  \footnotesize{lower bounds}\\
  \rule[2pt]{5pt}{0.5pt}\rule[2pt]{5pt}{0pt}\rule[2pt]{5pt}{0.5pt}\> 
    \footnotesize{upper bounds} \\
  \end{tabbing}
  \end{minipage}
}
\caption[Numerical data for $P_\conndb$ compared with the analytical results]
{Numerical data for $P_\conndb$ compared with the analytical results.
Parameters are set to $n = \sqrt{\ell}$ and $r = k \sqrt{\ell} \log_2 \ell$,
with different values of $k$. The numerical data 
was taken from \cite[Fig.~3]{SBV:range_assignment}.
} \label{SantiCompareFig}
\end{figure}

Santi and Blough also presented numerical data for $P_\conndb$, derived
by statistical simulation. In fact, 
their numerical method amounts to a Monte Carlo approximation of the integral
\begin{equation}
P_\conndb = \intcube d^nx \; \chi_\conndb(\xv).
\end{equation}
We will use their results from \cite[Fig.~3]{SBV:range_assignment} for comparison
with our analytical results. (The data given in \cite[Fig.~2]{SanBlo:transmitting_range}
is not much suited for our comparison, since most data points correspond to very
low values of $n$ or $\rho$.) Figure~\ref{SantiCompareFig} shows these data series
together with the upper and lower bounds from Lemma~\ref{dbEstimateLemm}, where
$P_\connpb$ is given by Theorem~\ref{pdiscPeriodicThm}. 
The numerical data is
compatible with the analytical bounds within the level of precision that would be expected
from a Monte Carlo type of approximation; since the underlying mathematical model
is identical in both cases, any differences can only be due to numerical precision.

In fact, Fig.~\ref{SantiCompareFig} suggests that the exact values of $P_\conndb$
are much nearer to our upper bounds than to the lower bound $P_\connpb$. This can
be understood at least on a heuristic level: In a situation where 
the largest part of $\Omega_n$ (more than 80\% of probability) corresponds to connected
networks, it would be expected that most of the remaining parts of 
$\Omega_n$ fall into $M_\discpb{1}$, and that $M_\discpb{2}$ and higher disconnected
events can rather be neglected. (Note that according to Theorem~\ref{pconnPbAsympThm}, the
$P_\discpb{k}$ asymptotically follow a Poisson distribution.) Thus,
the estimate in Eq.~\eqref{allKEstimate}, where we replaced
$P(M_\discpb{1} \cap S)$ with $P(\event{S})$, is quite tight in this case.

\chapswitch
\chapter{Quality Measures} \label{QualityChap}

When concerned with the structure of a MANET on a low level,
i.e. related to the mere connectivity of nodes,
the question arises by which quantitative properties the quality
of the network should be described.
Several measures have been proposed in the literature to that end.
 Most commonly, it is required that the
MANET should be (strongly) connected with high probability; 
however, this requirement turns out to be quite strong, 
and so one may want to consider more general, in particular weaker,
measures. 

In this chapter, we will introduce a general notion of 
so-called quality parameters 
for MANETs, and show that detailed results for specific parameters
can be obtained at least in simple models. 
In particular, this allows us to discuss
the scalability of such quality measures for large systems.

We will first define more exactly what a quality parameter is in our context,
and introduce a classification of such parameters according 
to their scaling behaviour;
this is done in Sec.~\ref{QparamGeneralSec}. We then consider several
specific quality parameters in Sec.~\ref{QparamCalcSec} and 
calculate their expectation value in the 1-dimensional MANET model. 
In Sec.~\ref{SkywaysSec}, we will compare our results to numerical
simulations conducted by Roth \cite{Rot:critical_mass} in a similar model.
Lastly, in Sec.~\ref{quantitSec}, give some examples of
quantitative predictions for MANET design that follow from our analysis.

\section{General properties of quality parameters} \label{QparamGeneralSec}

A quality parameter for MANETs in our context is a random variable
$Q : \Omega_n \to \rbb$, or rather a family of such random variables (for different
parameter values). The average ``quality'' of the MANET is then described
by its expectation value $\bar Q = \expect{Q}$. We will usually choose the range
of $Q$ to be $[0,1]$; however, this is only a matter of convention.

The definition of specific quality parameters naturally is very dependent
on the usage scenario and application. However, there is one overall property
that we wish to discuss in a general context: It relates to the scaling behaviour
of the system, since we are usually interested in the limit of large MANETs ($n \to \infty$).

Let us consider the $1$-dimensional MANET model from Chap.~\ref{ConnectChap}
for concreteness. If the quality parameter $Q$ is an ``intrinsic property''
of the system, that is related to its behaviour in the bulk, then one might expect
the following: If we take, say, two MANETs with identical parameters $n$, $r$, $\ell$,
and couple them together -- i.e., we join the two intervals and consider them
as a single network with the double node number, 
allowing connections between the two parts --,  
and if the original MANETs had a quality of $\bar Q = \expect{Q}$, 
then the joint MANET should have the same quality value $\bar Q$,
at least approximately for large systems. This would mean
\begin{equation}
	\expect{Q^{(n,r,\ell)}} \approx  \expect{Q^{(2n,r,2\ell)}}
	\text{ \quad or, equivalently, \quad}
	\expect{Q^{(n,\rho)}} \approx  \expect{Q^{(2n,\rho/2)}},
\end{equation}
referring to the normalized radio range. Of course, the same heuristic
argumentation should hold when tripling the system size, dividing it into parts, etc.; 
more generally,
the quality value should depend on $n \rho$ only, rather than on $n$ and $\rho$ 
independently. Let us formulate this more precisely.

\begin{defn} \label{intensiveDefn}
In the model of a 1-dimensional MANET, a family of random variables
$Q^{(n,\rho)}$ is called \emph{intensive}\footnote{
  The usage of the word \emph{intensive} is motivated by an analogy to 
  statistical physics: Here a thermodynamic variable, e.g. 
  a state parameter for a gas, 
  is called intensive
  if it does not change when the system is divided into parts; 
  examples include temperature, pressure, and particle density.
} if there exists a function
$\tilde Q: \rbb^+ \to \rbb$ with the following properties:
\begin{itemize}
  \item $\tilde Q$ is not globally constant;
  \item Given $\nu \in \rbb^+$ and a sequence $(\rho_n)$ in $\rbb^+$ such that 
  	    $n \rho_n \to \nu$ as $n \to \infty$, one has
  	    \begin{equation*}
  	      \expect{Q^{(n,\rho_n)}} \to \tilde Q (\nu).
  	    \end{equation*}
  	    
\end{itemize}
\end{defn}

Here the first condition is introduced in order to exclude ``trivial'' 
intensive parameters, such as those where always $\bar Q \to 0$ when 
$n \rho \to const$. Note that the parameter $n \rho = n r / \ell$ can be interpreted
as the ``non-statistical degree of coverage'' of the network: E.g. $n\rho = 1$ means that 
the radio range of all nodes combined covers the interval $[0,\ell]$
exactly once.

By the above definition, we do not mean to say that only intensive quality parameters
are relevant for our system, or that non-intensive parameters are not meaningful.
In fact, such non-intensive quality parameters may be required for some applications. 
However, one should keep in mind that these parameters may not scale well
for large systems: For example, if we need $n \rho \to \infty$ in order to keep the quality
level of the system constant as $n \to \infty$, then this means that
the average number of nodes per interval of length $r$  needs to grow
arbitrarily in the limit; thus we are 
likely to run out of local channel capacity. Hence applications
which rely on a high quality level with respect to non-intensive parameters
may not be feasible in networks with a high node number.

\section{Specific quality parameters}
\label{QparamCalcSec}
 
We will now investigate a number of specific quality parameters
and calculate their expectation value in the 1-dimensional
MANET model introduced in Chap.~\ref{ConnectChap}, where we will
always refer to the case of periodic boundary conditions.
Our choice of quality parameters mainly follows a discussion 
by Roth \cite{Rot:critical_mass}, who introduced four such
measures (segmentation, area coverage, vulnerability, and
reachability) in the context of a numerical simulation.
 
\subsection{Connectedness} \label{QconnSec}

One obvious choice for a quality parameter is the probability
that the network is connected, which we had already
investigated in Chap.~\ref{ConnectChap}. So, more formally, we set
$Q_\text{Connectedness} = \chi_\connpb$, where we know from 
Theorems~\ref{pconnPbAsympThm} and~\ref{pconnPbExtThm} that 
\begin{align} \label{connectedRes}
  \expect{Q_\text{Connectedness}} = P_\connpb  \quad&\to \exp(-e^{-\eta})
  \quad & \text { as } n \rho - \ln n \to \eta,
  \notag
  \\
   &\to 0
  \quad &  \text { as } n \rho \to \nu.
\end{align}
Thus $Q_\text{Connectedness}$ is \emph{not} an intensive parameter.
As discussed above, 
this means that applications relying on connectedness of the network
will not scale well in large systems. 

Closely related to connectedness is the quality measure of \emph{coveredness,}
investigated by Piret \cite{Pir:radio_connectivity} in 1-dimensional systems. 
Coveredness (not to be confused
with the \emph{area coverage} parameter that we will discuss in the next section)
measures whether each point in the interval $[0,\ell]$ is covered by the range of
at least one MANET node. It is clear that
we need precisely $y_i < 2 \rho$ for each next-neighbour distance $y_i$
to achieve that the interval
is completely covered, while the criterion for connectedness is $y_i < \rho$.
Thus, coveredness is related to connectivity by
\begin{equation}
  Q_\text{Coveredness}^{(n,\rho)}  =   Q_\text{Connectedness}^{(n,2\rho)}\; ,
\end{equation}
and we can apply the above result~\eqref{connectedRes} accordingly.

\subsection{Area coverage}

\emph{Area coverage} is the area $A_\text{covered}$ covered 
by the range of at least one MANET node, divided by the total
area $A_\text{total}$ of the system:
\begin{equation}
  Q_\text{Coverage} = \frac{A_\text{covered}}{A_\text{total}}.
\end{equation}
Its expectation value may be understood as the probability that an external
network node, with its position randomly chosen, will be able to connect
to at least one of the $n$ nodes of the MANET.

In our 1-dimensional model, ``area'' is to be understood as the length 
of the corresponding line segments. Note that through dividing by $A_\emph{total} = \ell$,
our parameter $Q_\text{Coverage}$ is scaling in the sense of 
Definition~\ref{scalingDef}; thus we may again pass to the normalized radio range
and set $A_\emph{total}= 1$. It is also easy to express  $Q_\text{Coverage}$
in terms of next-neighbour variables: The distance $y_i$ leaves an area uncovered
if $y_i > 2\rho$; if so, the length of that area is $y_i-2\rho$. Thus we get
the following expression for $Q_\text{Coverage}$:
\begin{equation} \label{coverageSum}
	Q_\text{Coverage} = 1 - \sum_{i=1}^n (y_i-2\rho) \; \theta(y_i-2\rho).
\end{equation}
In order to determine its expectation value, we will calculate 
\begin{equation}
	\expect{  (y_i-2\rho) \; \theta(y_i-2\rho) } = \intcube  d\muEqualT{n} (\yv) \;
	  (y_i-2\rho) \; \theta(y_i-2\rho)
\end{equation}
for each fixed $i$, where we will assume $\rho < 1/2$  (otherwise, we trivially
have $\bar Q_\text{Coverage}=1$). Lemma~\ref{tnScalingLemm} 
and Proposition~\ref{tnEqualProp} of Appendix~\ref{SimplexApp}
then yield
\begin{equation}
	\expect{  (y_i-2\rho) \; \theta(y_i-2\rho) } 
	= (1-2\rho)^{n-1} \intcube  d\muEqualT{n}(\yv)\; (1-2\rho) y_i 
	= \frac{1}{n}(1-2\rho)^n 
\end{equation}
Inserting into the expectation value of  \eqref{coverageSum},
we obtain
\begin{equation}
   \expect{Q_\text{Coverage}} = 1 - (1-2\rho)^n.
\end{equation}
(Again, this is valid for $\rho < \frac{1}{2}$.) Using Taylor approximation
$\ln (1-x) = -x + O(x^2)$, we have
\begin{equation} \label{logApproxNQ}
   \ln \; (1-2\rho)^n = -2n\rho + O(n\rho^2);
\end{equation}
thus, in the limit $n\rho \to \nu$ (where $n \to \infty$, $\rho \to 0$, and
$n\rho^2 \to 0$), the area coverage converges to
\begin{equation} \label{coverageResult1d}
   \expect{Q_\text{Coverage}} \to 1 - e^{-2\nu}. 
\end{equation}
This means that the area coverage is an intensive quality parameter.

\subsection{Segmentation} \label{SegmentationSec}

The \emph{segmentation} of a MANET counts the number of 
disconnected segments in the network, i.e. the number of subgraphs
into which the network graph is separated: We set
\begin{equation} \label{segmentationDef}
  Q_\text{Segmentation} = \frac{ \text{\# of network segments} }{ \text{\# of network nodes} }.
\end{equation}
In order to take account of the periodic boundary conditions, we will count
the strongly connected situation (the event $\conndb$) as having
0 network segments. (This explains the slightly modified setting
in Eq.~\eqref{segmentationDef} when compared with the original definition
by Roth~\cite{Rot:critical_mass}, who defined 
\begin{equation} \label{segmentationDefRoth}
  Q_\text{Segmentation} = \frac{ \text{\# of network segments} -1}{ \text{\# of network nodes} -1}.
\end{equation}
This difference is rather a matter of convenience and should not play a role
in the limit of large systems.) 

Within our 1-dimensional system, it is easy to derive an explicit expression
for the segmentation: We know that the event $\discpb{k}$ corresponds to 
a situation with exactly $k$ network segments. Since these events are disjoint,
and since their union (over $k = 0 \ldots n$) exhausts the sample space $\Omega_n$,
it follows that
\begin{equation}
  Q_\text{Segmentation} = \frac{1}{n} \sum_{k=0}^n k \chi_\discpb{k}
\end{equation}
and consequently
\begin{equation}
  \expect{Q_\text{Segmentation}} 
    = \frac{1}{n} \sum_{k=0}^n k P_\discpb{k}.
\end{equation}
The probabilities under the sum are known from Theorem~\ref{pdiscPeriodicThm}:
\begin{equation}
  \expect{Q_\text{Segmentation}} 
    = \frac{1}{n} \sum_{k=0}^n \sum_{j=k}^{[1/\rho]} (-1)^{j-k} 
    	k \binom{j}{k} \binom{n}{j} (1-j\rho)^{n-1}.
\end{equation}
Now observe that in the sum over $j$, we may as well replace the lower limit with $0$, 
since the binomial coefficient $\binom{j}{k}$ vanishes for $j<k$.
We may then exchange the order of summation and get
\begin{equation} \label{segmSumExchanged}
  \expect{Q_\text{Segmentation}} 
    = \frac{1}{n}  \sum_{j=0}^{[1/\rho]} 
     (-1)^{j}  \binom{n}{j} (1-j\rho)^{n-1}
    	\sum_{k=0}^n (-1)^k k \binom{j}{k}.
\end{equation}
Likewise, we may replace $n$ with $j$ in the upper limit of the sum over $k$,
since the summand vanishes for $k>j$ as well as for $j>n$ due to the binomial factors.
Referring to Lemma~\ref{kDeltaSumLemm} in Appendix~\ref{SummationApp}, 
we know that
\begin{equation}
  \sum_{k=0}^j (-1)^k k \binom{j}{k} = 
  \begin{cases}
  	 -1 & \text{if } j=1, \\
  	 0 & \text{otherwise}.
  \end{cases}
\end{equation}
So in Eq.~\eqref{segmSumExchanged}, only the summand for $j=1$ remains. 
Assuming $\rho < 1$, that leads to the result
\begin{equation} \label{segmentResult}
  \expect{Q_\text{Segmentation}} = (1-\rho)^{n-1}.
\end{equation}
With arguments as in Eq.~\eqref{logApproxNQ}, this means that in the limit $n\rho \to \nu$,
\begin{equation}
    \expect{Q_\text{Segmentation}}  \to e^{-\nu},
\end{equation}
so $Q_\text{Segmentation}$ is an intensive parameter as well.

\subsection{Vulnerability} \label{VulnSec}

The next quality parameter we will consider is related to the question 
how much the network quality or topology changes when a single
node is removed from the network. Specifically, we define the
\emph{importance} of the network node with number $j$ as
\begin{equation}
  I_j := \max \{0, (\text{\# segments with node $j$ removed}) 
  	- (\text{\# segments}) \};
\end{equation}
i.e. $I_j$ is the number of network segments which are created
by switching off node $j$ in the current configuration.
Nodes with $I_j > 0$ make the network ``vulnerable'' against changes.
This motivates to define the \emph{vulnerability} of the network as
\begin{equation} \label{vulnDef}
  Q_{\text{Vulnerability}} = \frac{1}{n} \sum_j I_j.
\end{equation}

In our 1-dimensional model, the importance of a node is
either 1 (if removing the nodes splits the respective 
network segment in two) or 0. The ordering of nodes is not
of relevance for Eq.~\eqref{vulnDef}; so we may describe the
event $j\event{-IMPORTANT}$ (meaning that $I_j=1$) 
directly in next-neighbour coordinates as
\begin{equation}
  M_{j\event{-IMPORTANT}} = \{ \yv   \,|\, (y_{j-1}  < \rho) \wedge (y_{j} < \rho) \wedge
  							(y_{j-1}+y_j \geq \rho) \},
\end{equation}
where the coordinate indices are understood ``modulo $n$,'' i.e. $y_{0}$ is identified
with $y_n$. We will assume $n \geq 2$ in the following, 
so that $y_j$ and $y_{j-1}$ are independent coordinates. 
Taking the complement of the set above, we can say that
\begin{equation}
  P_{j\event{-IMPORTANT}} = 1 - P(M_{j\event{-IMPORTANT}}^c)
  = 1 - P( 
  y_{j-1}  \geq \rho \vee y_{j} \geq \rho \vee
  							y_{j-1}+y_j <\rho).
\end{equation}
On the last expression, we apply the inclusion-exclusion formula from 
Appendix~\ref{IncExcApp}; this yields\footnote{
More specifically, we apply Theorem~\ref{incExcThm} with 
respect to the event $C_1$ and for $n=3$ (with notation as in the theorem).
}
\begin{align} \label{pjImportant}
  P_{j\event{-IMPORTANT}} = \;
  & 1  - P(y_{j-1} \geq \rho)  - P(y_j \geq \rho)   - P(y_{j-1}+y_j < \rho) 
 + P( y_j \geq \rho \wedge y_{j-1} \geq \rho) 
   \notag\\
  &+ P( y_{j-1} \geq \rho \wedge y_{j-1}+y_j < \rho)
  + P( y_j \geq \rho \wedge y_{j-1}+y_j < \rho)
  \notag\\
  & - P( y_{j-1} \geq \rho \wedge y_j \geq \rho \wedge y_{j-1}+y_j < \rho).
\end{align}
The last three summands of this expression obviously vanish.
Moreover, we know from Lemma~\ref{pGeqLemm} that 
\begin{align}
  &P(y_j \geq \rho) = P(y_{j-1} \geq \rho) = (1-\rho)^{n-1},
  \\
  &P(y_{j-1} \geq \rho \wedge y_{j} \geq \rho) = (1-2\rho)^{n-1};
\end{align}
here we have assumed $\rho < 1/2$. 
Further, Lemma~\ref{tnDoubleThetaLemm} in Appendix~\ref{SimplexApp}
shows that
\begin{multline} \label{doubleTheta}
  P(y_{j-1} + y_j < \rho) 
  = \intcube d\muEqualT{n} (\yv)\; \theta(\rho-y_{j-1}-y_j)
  \\
  = 1 - (1-\rho)^{n-2} (1+(n-2)\rho).
\end{multline}
Combining Eqs.~\eqref{pjImportant} to~\eqref{doubleTheta},
we have shown that
\begin{equation}
  P_{j\event{-IMPORTANT}} = (n\rho-1)(1-\rho)^{n-2} + (1-2\rho)^{n-1}.
\end{equation}
Inserting into Eq.~\eqref{vulnDef}, we have obtained that for
$n \geq 2$ and $\rho < 1/2$:
\begin{equation}
 \expect{Q_\text{Vulnerability}}
   = \frac{1}{n} \sum_{j=1}^{n} P_{j\event{-IMPORTANT}} 
   = (n\rho-1)(1-\rho)^{n-2} + (1-2\rho)^{n-1}.
\end{equation}
A Taylor approximation (as in the previous sections) then leads
us to the following asymptotic behaviour in the limit $n\rho \to \nu$:
\begin{equation}
 \expect{Q_\text{Vulnerability}} \to
   (\nu -1) e^{-\nu} + e^{-2\nu}.
\end{equation}
Thus, the vulnerability is an intensive quality parameter as well.

\subsection{Reachability}

The \emph{reachability} parameter is concerned
with the number of nodes that can be reached from a given node
(in a multi-hop fashion), or, alternatively speaking, with the size
of the segments of the network. We define the 
reachability of some fixed node $j$ as
\begin{equation}
  R_j := \frac{ \text{\# of nodes reachable from node $j$} }{n}.
\end{equation}
Here we do not count the node itself as reachable, unless the network is 
strongly connected (i.e. the node can ``reach itself'' via the boundary).
We define our quality parameter, the average reachability, as
\begin{equation}
Q_\text{Reachability} = \frac{1}{n}  \sum_{j=1}^{n} R_j.
\end{equation}
Again, we have introduced a slight difference 
compared to the original definition by Roth \cite{Rot:critical_mass}
which accounts for the periodic boundary conditions and vanishes for $n \to \infty$.
Following our above discussion, the value of $Q_\text{Reachability}$ is
\begin{itemize}
\item 
  1 in the event $\connpb$,
\item
  $(n-1)/n$ in the event $\discpb{1}$,
\item
  more generally, $n^{-2} \sum_{i=1}^k b_i(b_i-1)$
  in the event $\discpb{k}$, $k \geq 1$, where $b_i$ are the sizes of the
  $k$ network segments.
\end{itemize}
To get a more explicit description of the latter case for $k \geq 2$, 
we define the events
$\segm{m}{b}$, where $m \in \{1,\ldots,n\}$, $b \in \{1,\ldots,n-1\}$,
which describe that a segment of the network begins exactly
at node $m$, extending ``to the right,'' and has a size of exactly $b$ nodes. 
(The node indices are counted in sorted coordinates, and are defined modulo $n$.)
This can be formally expressed as
\begin{equation} \label{chiSegmDef}
  \chi_\segm{m}{b} (\yv) = \theta(y_{m-1}-\rho) \; \theta (y_{m+b-1}-\rho) \;
    \prod_{i=m}^{m+b-2} \theta(\rho-y_i).
\end{equation}
It is then easy to sum over the size of the segments:
Since the events $\segm{m}{b}$ are obviously disjoint from $\connpb$ and $\discpb{1}$,
one simply has
\begin{equation} \label{reachSum}
Q_\text{Reachability} =
  \chi_\connpb
  +  \frac{n-1}{n} \chi_\discpb{1}
  + \sum_{m=1}^{n} \sum_{b=1}^{n-1} \frac{b(b-1)}{n^2} \chi_\segm{m}{b}.
\end{equation}
Since the expectation value of the first two summands has already been calculated
in Chap.~\ref{ConnectChap}, it only remains to calculate $P_\segm{m}{b}$ in order to
determine $\expect{Q_\text{Reachability}}$.
Using the definition in Eq.~\eqref{chiSegmDef}, and applying Lemma~\ref{tnScalingLemm}
twice, we see that for $n \geq 2$ and $ \rho < 1/2$,
\begin{align}
P_\segm{m}{b} &=
  \intcube d\muEqualT{n}(\yv) \;
  \theta(y_{m-1}-\rho) \; \theta (y_{m+b-1}-\rho) \;  \prod_{i=m}^{m+b-2} \theta(\rho-y_i)
  \notag\\
  &=
  (1-\rho)^{n-1}
  \intcube d\muEqualT{n} (\yv) \;
  \theta (y_{m+b-1}- \frac{\rho}{1-\rho}) \;  \prod_{i=m}^{m+b-2} \theta(\frac{\rho}{1-\rho}-y_i)
  \notag\\
  &=
  \underbrace{(1-\rho)^{n-1} (1-\frac{\rho}{1-\rho})^{n-1}}_{ =(1-2\rho)^{n-1}}
  \intcube d\muEqualT{n} (\yv) 
   \prod_{i=m}^{m+b-2} \theta(\frac{\rho}{1-2\rho}-y_i)
  \notag\\
  &=
  (1-2\rho)^{n-1} 
   P(y_m < \rho' \wedge \ldots \wedge y_{m+b-2} < \rho'), 
\end{align}
where $\rho' = \rho/(1-2\rho)$.
For determining the probabilities $P(y_m < \rho' \wedge \ldots)$, we once again
use the inclusion-exclusion formula\footnote{
More precisely, we use Theorem~\ref{incExcThm} with respect to the event $C_1$
and with $(b-1)$ in the place of $n$.
}
 of Appendix~\ref{IncExcApp}:
\begin{multline}
   P(y_m < \rho' \wedge \ldots \wedge y_{m+b-2} < \rho')
   = 1 - P(y_m \geq \rho' \vee \ldots \vee y_{m+b-2} \geq \rho')
   \\
   = 1 - \sum_{j=1}^{b-1} (-1)^{j-1} \binom{j-1}{0} S_j
   = \sum_{j=0}^{b-1} (-1)^j S_j,   
\end{multline}
where
\begin{equation}
  S_j  = \sum_{ \{m_1,\ldots,m_j\} \subset\{ m,\ldots,m+b-2\} }
   P(y_{m_1} \geq \rho' \wedge \ldots \wedge y_{m_j} \geq \rho').
\end{equation}
We already know the probability under the sum by Lemma~\ref{pGeqLemm}.
Applying this result leads us to
\begin{equation} \label{pbOrig}
   P(y_m < \rho' \wedge \ldots \wedge y_{m+b-2} < \rho')
   = \sum_{j=0}^{[1/\rho']} (-1)^j \binom{b-1}{j} (1-j\rho')^{n-1}.
\end{equation}
Now we can assemble our results,
together with the expressions for $P_\connpb$ and $P_\discpb{1}$
from Theorem~\ref{pdiscPeriodicThm}, in order to  
determine the expectation value of Eq.~\eqref{reachSum}:
This gives
\begin{align} \label{reachResultFin}
 \expect{Q_\text{Reachability}}
=&  
  P_\connpb
  +  \frac{n-1}{n} P_\discpb{1}
  + \sum_{m=1}^{n} \sum_{b=1}^{n-1} \frac{b(b-1)}{n^2} P_\segm{m}{b}
\notag\\
=&  \sum_{j=0}^{[1/\rho]} (-1)^j \binom{n}{j} (1-j\rho)^{n-1}
  +  \frac{n-1}{n} \sum_{j=0}^{[1/\rho]} (-1)^j j \binom{n}{j} (1-j\rho)^{n-1}
\notag \\
  &+ n^2 (1-2\rho)^{n-1} \sum_{b=1}^{n-1} \frac{b(b-1)}{n^3} 
    \sum_{j=0}^{[1/\rho']} (-1)^j \binom{b-1}{j} (1-j\rho')^{n-1},
\end{align}
where $\rho' = \rho/(1-2\rho)$, and we assume $n\geq 2$, $\rho < 1/2$.

While this explicit expression is rather complicated, we can derive
a much simpler result for the limit $n \to \infty$, where we 
consider $n\rho-\ln n \to \eta$ as in Sec.~\ref{QconnSec}.
We already know the limit values of $P_\connpb$ and $P_\discpb{1}$
from Theorem~\ref{pconnPbAsympThm}. It is also easy to see that
\begin{equation}
  \ln (n^2 (1-2\rho)^{n-1}) = 2 \ln n - 2 n \rho + O(\rho) + O(n\rho^2) \to -2 \eta,
\end{equation}
so the factor $n^2 (1-2\rho)^{n-1}$ converges to $e^{-2\eta}$. It remains
to determine the asymptotic behaviour of the sum over $b$. 
The idea here is to understand the sum (for large $n$) as the approximation
of a Riemann integral, where the integration variable $\beta = b/n$ 
ranges from $0$ to $1$. Since the calculation is somewhat involved,
we state it as a separate lemma.

\begin{lemm} \label{reachAsympLemm}
Let $\eta \in \rbb$, $(\rho_n) \subset \rbb^+$ such that 
$n \rho_n - \ln n \to \eta$ as $n \to \infty$, and let
$\rho_n' := \rho_n / (1-2\rho_n)$. Then one has
\begin{equation*}
  \sum_{b=1}^{n-1} \frac{b(b-1)}{n^3} 
    \sum_{j=0}^{[1/\rho_n']} (-1)^j \binom{b-1}{j} (1-j\rho_n')^{n-1}
    \longnlim
    \int_{0}^1 d\beta \;\beta^2 \exp(-\beta e^{-\eta}).
\end{equation*}
\end{lemm}

\begin{proof}
In the following, we keep $\eta$ fixed and set
\begin{align}
  f(\beta) &= \exp(-\beta e^{-\eta}), \\
  f_n(\beta) &= \sum_{j=0}^{[1/\rho_n']} (-1)^j \binom{[n\beta]-1}{j} (1-j\rho_n')^{n-1}, \\
  \quad \text{and} \qquad
  a_n &= j! \binom{b-1}{j} (1-j\rho_n')^{n-1}.
\end{align}
We obviously have $|f(\beta)| \leq 1$ for $\beta \in [0,1]$, 
and we also know that  $|f_n(\beta)| \leq 1$ for 
$\beta = b/n$, $b \in \{1,\ldots,n\}$, 
since the $f_n(b/n)$ are defined as probabilities (cf. Eq.~\eqref{pbOrig};
we can easily extend this to the case $b=n$).
This is useful for simplifying the proposition of the lemma: Since
\begin{equation}
  | \frac{n(n-1)}{n^3} f_n(1)| \leq \frac{1}{n} \to 0
\end{equation}
and
\begin{equation}
  | \sum_{b=1}^n \frac{b^2-b(b-1)}{n^3} f_n(b/n)| \leq \frac{1}{n^2} 
  \sum_{b=1}^n \frac{b}{n} \leq \frac{1}{n} \to 0,
\end{equation}
we can equivalently prove that
\begin{equation}
 \big|  \sum_{b=1}^{n} \frac{b^2}{n^3} f_n\big(\frac{b}{n}\big)
    -
    \int_{0}^1 d\beta \;\beta^2 f(\beta) \big|
    \longnlim 0.
\end{equation}
However, since $f$ is integrable, it is clear by the definition of
the Riemann integral that
\begin{equation}
 \big|  \sum_{b=1}^{n} \frac{1}{n} \;\frac{b^2}{n^2} 
 f\big(\frac{b}{n}\big)
    -
    \int_{0}^1 d\beta \;\beta^2 f(\beta) \big|
    \longnlim 0.
\end{equation}
Thus, it only remains to verify that
\begin{equation} \label{estimateGoal}
   \sum_{b=1}^{n} \frac{b^2}{n^3} 
   \big|  f_n\big(\frac{b}{n}\big) - f\big(\frac{b}{n}\big)  \big|
    \longnlim 0.
\end{equation}
To that end, we need an estimate of $|f_n(b/n)-f(b/n)|$ that is uniform 
in $b$. We will construct this estimate by refining the methods
developed in the proof of Theorem~\ref{pconnPbAsympThm}, using
notation as introduced there.\footnote{
Note that the parameter $k$ in Theorem~\ref{pconnPbAsympThm}
must be set to $0$ for our purposes.
}

Regarding the terms $a_j$, we can certainly say that for $j \leq b-1$,
\begin{equation}
  j! \binom{b-1}{j} = \frac{(b-1)!}{(b-1-j)!} \leq (b-1)^j \leq n^j,
\end{equation}
independent of $b$; the same is true for $j > b-1$ (where the binomial 
coefficient vanishes). We can then apply the same construction
that lead to Eq.~\eqref{ajRemainder}. Thus, for given $\epsilon>0$, 
we can find $j_0$ and $n_0$ such that for any $n \geq n_0$,
\begin{equation} \label{fAjRemainder}
  \big| \sum_{j=j_0}^{ [1/\rho_n]} \frac{(-1)^j}{j!} a_j  \big| \leq 2 \epsilon.
\end{equation}
and at the same time, for any $b \in \{1,\ldots,n\}$,
\begin{equation} \label{fExpRemainder}
  \big| \sum_{j=j_0}^{ \infty } \frac{(-1)^j}{j!} \big(\frac{b}{n} e^{-\eta}\big)^j  \big|
  \leq \epsilon .
\end{equation}
(Note that we can find such an estimate independent of $b$, since
the power series $\sum_j x^j/j!$ converges uniformly on the interval $[-e^{-\eta},0]$.)

Now it remains to handle the terms for $j < j_0$; we have to find a
uniform estimate for $| (\frac{b}{n}e^{-\eta})^j-a_j|$ for all $b$ at fixed $j$.
Let us first consider those terms where $b \geq \epsilon n$,
where we can assume that $j_0 < \epsilon n$ (possibly after increasing $n_0$).
We know that
\begin{equation}
  a_j / (\frac{b}{n}e^{-\eta})^j = \frac{(b-1)!}{(b-1-j)!} \frac{1}{b^j} \;
     n^j e^{\eta j} (1-j\rho_n')^{n-1}.
\end{equation}
Only the first factors in this expression depend on $b$; they are
\begin{equation}
   \frac{(b-1)!}{(b-1-j)!} \frac{1}{b^j}  
   = \frac{b-1}{b}  \; \ldots \frac{b-j}{b}.
\end{equation}
Each of the factors of the form $(b-i)/b$ converges to 1, more explicitly:
\begin{equation}
   |  \frac{b-i}{b} -1  | = \frac{i}{b} \leq \frac{j_0}{ \epsilon n}.
\end{equation}
Thus we can control the convergence of these factors independent of $b$
(with $j_0$ still being fixed). Moreover, we find 
-- just as in Eq.~\eqref{secondFactorConv} --
that 
\begin{equation}
   n^j e^{\eta j} (1-j\rho_n')^{n-1} \to 1,
\end{equation}
where the term does not depend on $b$. Thus the convergence of 
$a_j / (\frac{b}{n}e^{-\eta})^j \to 1$ is uniform in $b$, given that
$b \geq \epsilon n$. Summarizing this with Eqs.~\eqref{fAjRemainder} 
and~\eqref{fExpRemainder}, we have found that
\begin{equation}
  \forall \epsilon > 0 \; \exists n_1 \; \forall n \geq n_1\;
  \forall b \in \{[\epsilon n]+1,\ldots,n\}:
  \big| f_n\big( \frac{b}{n} \big) -f\big( \frac{b}{n} \big)  \big| < 4 \epsilon.
\end{equation}
For $b \leq \epsilon n$, we will use the rough estimate
\begin{equation}
   \big| f_n\big( \frac{b}{n} \big) -f\big( \frac{b}{n} \big)  \big| \leq 2.
\end{equation}
Now combining these bounds, we can establish Eq.~\eqref{estimateGoal}:
For $n \geq n_1$, we have
\begin{equation}
  \sum_{b=1}^n \frac{b^2}{n^3} 
   \big| f_n\big( \frac{b}{n} \big) -f\big( \frac{b}{n} \big)  \big| 
   \leq 4 \epsilon   \sum_{b=[\epsilon n]+1}^n \frac{b^2}{n^3} 
      \;+ \; 2   \sum_{b=1}^{[\epsilon n]} \frac{b^2}{n^3} 
   \leq 4 \epsilon \frac{1}{n} n + 2 \frac{1}{n} [\epsilon n]
   \leq 6 \epsilon.
\end{equation}
This finally proves Eq.~\eqref{estimateGoal} and hence the lemma.
\end{proof}

Of course, the integral that we established as a limit value in the above lemma
is easy to solve (twice integrating by parts): One has
\begin{multline}
  \int_{0}^1 d\beta \;\beta^2 \exp(-\beta e^{-\eta})
   =  -\exp(-\beta e^{-\eta}) (2 e^{3 \eta} + 2 e^{2 \eta} \beta + e^{\eta} \beta^2)   \Big|_0^1
   \\
   = -\exp(-e^{-\eta}) (e^\eta + 2 e^{2\eta} + 2 e^{3\eta}) + 2 e^{3\eta}.
\end{multline}
Now collecting our results on $\expect{Q_\text{Reachability}}$ in Eq.~\eqref{reachResultFin},
where the limits for $P_\connpb$ and $P_\discpb{1}$ are 
known from Theorem~\ref{pconnPbAsympThm}, we can establish that
\begin{equation}
  \expect{Q_\text{Reachability}}  \to 2 e^\eta - (1+2 e^\eta )\exp(-e^{-\eta}) 
  \quad\text{as}\; n\rho-\ln n \to \eta.
\end{equation}
By a monotony argument similar to the one which lead to Theorem~\ref{pconnPbExtThm},
we can show that 
$\expect{Q_\text{Reachability}} \to 0$ as $n \rho \to \nu$; 
so the reachability is \emph{not} intensive.

\newcommand{\PBS}[1]{\let\temp=\\#1\let\\=\temp}

\begin{table}
\begin{tabular}{p{0.15\textwidth}|p{0.30\textwidth}|p{0.30\textwidth}|c}
parameter
&
\multicolumn{2}{|c|}{expectation value}
&
intensive?
\\
&
\PBS\centering at finite $n \geq 2$, $\rho < 1/2$
&
\PBS\centering asymptotic
&
\\
\hline
$Q_\text{Connectedness}$
&
\begin{displaymath}
 \sum_{j=0}^{[1/\rho]} (-1)^j \binom{n}{j} (1-j\rho)^{n-1} 
\end{displaymath}
&
\,

$\exp(-e^{-\eta}) $ 

\quad as~$n\rho - \ln n \to \eta$
&
no
\\
\hline
$Q_\text{Coverage}$
&
$1 - (1-2\rho)^n$
&
%\begin{displaymath}
  $1 - e^{-2 \nu} $
  
  \quad as $n\rho \to \nu$
%\end{displaymath}
&
yes
\\
\hline
$Q_\text{Segmentation}$
&
$(1-\rho)^{n-1}$
&
  $e^{- \nu}$
  
  \quad as $n\rho \to \nu$
&
yes
\\
\hline
$Q_\text{Vulnerability}$
&
$(n\rho-1)(1-\rho)^{n-2}$ 

$\;+ (1-2\rho)^{n-1}$
&
  $(\nu-1)e^{- \nu} +e^{-2\nu} $
  
  \quad as $n\rho \to \nu$
&
yes
\\
\hline
$Q_\text{Reachability}$
&
see Eq.~\eqref{reachResultFin}
&
%\begin{displaymath}
  $2 e^\eta - (1+2e^\eta)\exp(-e^{-\eta})$
  
  \quad as~$n\rho-\ln n \to \eta$
%\end{displaymath}
&
no
\\
\hline
\end{tabular}
\caption{Overview of the results for quality parameters}
\label{qResultTable}
\end{table}

\newcommand{\skywaysNumLegend}{
\fbox{
  \begin{minipage}{0.15\textwidth}
  \begin{tabbing}
  \\
  \footnotesize{$\bullet$} \;\quad \= \footnotesize{simulation results}\\
  \rule[2pt]{15pt}{0.5pt} \>  \footnotesize{analytical results}\\
  \end{tabbing}
  \end{minipage}
}
}

\newcommand{\linlogComparison}[3]{

\begin{figure} 
\raggedleft
\skywaysNumLegend

\centering
\begin{minipage}{0.80\textwidth}
  \subfigure[linear]  {
     \includegraphics*[width=0.95\textwidth]{#2_linear.\figext}
  }
  \\
  \subfigure[logarithmic]  {
      \includegraphics*[width=0.95\textwidth]{#2_log.\figext}
  }
\end{minipage}

\caption{#3} \label{#1}
\end{figure}

}

\section{Comparison with simulations} \label{SkywaysSec}

We will now aim at comparing our results on quality parameters,
which are summarized in Table~\ref{qResultTable}, to the simulation
data obtained by Roth~\cite{Rot:critical_mass}. 

\begin{figure} 
\begin{center}
\includegraphics[height=0.3\textheight]{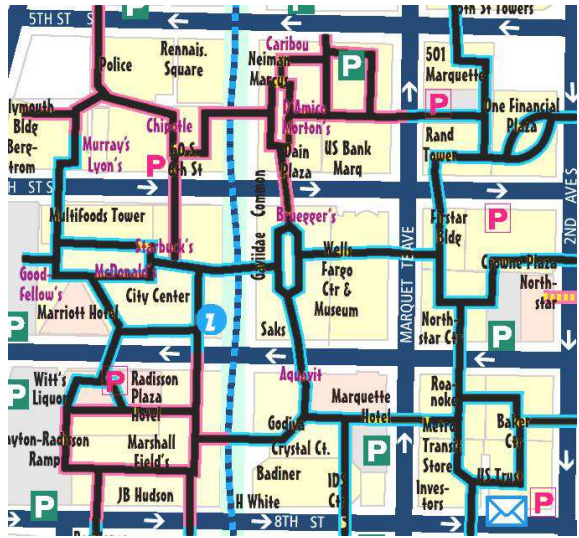}
\end{center}
\caption[Original map of the Skyways]
{Original map of the Skyways \cite{SkywaysSite}}
 \label{SkywaysOriMapFig}
\end{figure}

\begin{figure} 
\begin{center}
  \includegraphics*[height=0.4\textheight]{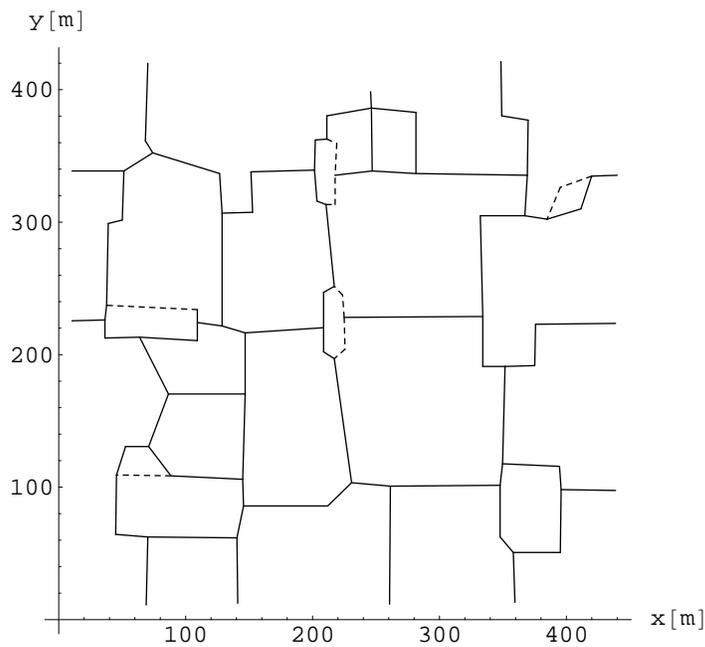}
\end{center}
\caption[Idealized map of the Skyways]
{Idealized map of the Skyways. Dashed line segments 
were not considered for determining the effective length $\ell$ (see text).} 
\label{SkywaysAbstMapFig}
\end{figure}

In contrast to the quite simplistic assumptions of our model,
Roth aimed at a more realistic network topology;
he chose part of the map of the Downtown Minneapolis shopping center
as the basis for his simulation (cf. Fig.~\ref{SkywaysOriMapFig}).
This shopping center consists of a number of towers which
are connected on the first floor via bridges, so-called ``Skyways'';
we consider users with wireless devices moving along these paths
(see Fig.~\ref{SkywaysAbstMapFig}).

This model is in a way quite similar to ours and largely makes
the same overall assumptions: Network nodes move independently at random 
on 1-dimensional paths; the radio range of all nodes is equal
with a sharp cutoff at radius $r$. However, there are a number of
important differences:

First, while we based our analysis on a static model
(assuming ergodicity for mobile nodes), Roth considered
an explicit motion model: Users move at constant speed along 
a line segment, and choose a new speed and direction once
they have reached the end of a segment. Certainly, one would
expect that this model also leads to an equal distribution
of nodes on the line segments in the long run; however, this
is not explicitly modelled.

Second, Roth considered a 2-dimensional radio propagation, 
in contrast to our 1-dimensional model; i.e. two nodes are
connected when their distance is smaller the $r$ on the \emph{plane}
rather than along the line segments. 
(No shielding by buildings, walls, etc. between the different
paths was taken into account.)
In most cases, this is equivalent
to our 1-dimensional propagation, since neighbouring line segments
are usually further than $r$ apart (cf. Fig.~\ref{SkywaysAbstMapFig}); 
however, there are some exceptions.
We will discuss this in more detail below.

Third, as already noted, the topology of the line segments
is much more complex than in our simplistic model, including
both open and closed curves.

Before we can compare our results to those of Roth,
we must first determine the parameters of our model 
that correspond to the situation considered by Roth. 
The radio
range was chosen as $r = 30 m$ (the indoor communication
range of IEEE 802.11b Wireless LAN), which we can directly
transfer to our situation. The system length $\ell$
is more difficult to determine: While it might seem obvious
to set $\ell$ as the total length of all line segments 
in the system (see Fig.~\ref{SkywaysAbstMapFig}), there are
two corrections we wish to make. These are due to the 2-dimensional
propagation model used by Roth.

On the one hand, Roth's model allows communication between
nodes on parallel (or nearly parallel) line segments whose
distance is less than the radio range. In our model, however, nodes
can only communicate in direction of the line segment.
Thus the range of a network node covers additional 
segment length in Roth's calculations,
the more the nearer such parallel line segments are located.
We will roughly accommodate this effect by the following procedure:
Whenever two parallel line segments in the map are 
not further than $r/2$ apart, we will only count one of them
for determining the total system length $\ell$. The line segments
that were left out due to this procedure are marked as dashed lines
in  Fig.~\ref{SkywaysAbstMapFig}.

On the other hand, there is another effect at those points
were at least 3 line segments meet. Due to the 2-dimensional propagation model,
nodes which are located near such a point can reach other nodes
in line segments of approximately $3r$ in length ($1r$ in each direction);
in our model from Chap.~\ref{ConnectChap}, however, nodes can only 
reach an ``area'' of $2r$ in length. In order to compensate this
difference, we will subtract $1r$ from the parameter $\ell$
for each such point on the map. There are 30 points of the mentioned
type on the map, not counting line segments that were left out
due to the procedure described earlier. This leaves us with an 
effective length of
\begin{equation}
  \ell = \unit{3363}{m} - 30 \cdot \unit{30}{m} = \unit{2463}{m}.
\end{equation}
Of course, these ``ad hoc corrections'' are only very rough 
and cannot be traced back directly to the statistical description.
They also do not account for all effects that relate to differences
between the models -- for example, the 2-dimensional radio propagation
certainly has an effect that relates to points where only 2 segments meet,
while the effect around the 3-segment points may have been over-estimated;
also, we do not account for the increased density of nodes in the areas
where two line segments run in parallel. However, we shall see that with
the corrections introduced, we can already get a good match between the 
results that the two models predict.

\linlogComparison{coverageCompareFig}{skyways_coverage}
{Analytical and simulation results for the area coverage parameter}

\linlogComparison{segmentationCompareFig}{skyways_segmentation}
{Analytical and simulation results for the segmentation parameter}

\linlogComparison{vulnCompareFig}{skyways_vulnerability}
{Analytical and simulation results for the vulnerability parameter}

After having fixed the parameters, let us now turn to a direct comparison
of the data. Roth did not consider connectedness as a quality parameter,
since in fact (as discussed above) strong connectivity would be a quite strict 
condition for networks of reasonable size. So we will discuss 
coverage, segmentation, vulnerability, and reachability. For all these parameters,
we will compare the numerical results of \cite{Rot:critical_mass}
with our explicit results listed in Table~\ref{qResultTable}, where we will use
the exact formulas rather than the asymptotic approximations.
(In most cases, the difference between the asymptotic approximation and exact value
is however so small that it would hardly be visible in the graphs.)

Let us start with the area coverage parameter, shown in Fig.~\ref{coverageCompareFig}.
The linear plot shows that both models nearly agree in absolute values for 
$n=50$ and $n=100$, and in the asymptotic behaviour as $n \to \infty$
(where both graphs approach 1), while there is some difference 
at medium values of $n$. However, the logarithmic plot reveals that
our 1-dimensional model systematically differs from Roth's simulation,
which shows a much lower area coverage at high $n$. An explanation for
this difference might be boundary effects in Roth's model: 
Possibly, some peripheral parts of the Skyways were not as densely 
covered with nodes as one would expect from the equal distribution.
Still, the absolute difference between the models is below 5\%, and
the models agree with respect to their qualitative behaviour.

The data for segmentation is shown in Fig.~\ref{segmentationCompareFig}.
It shows a good fit between the models, both on the linear and
logarithmic scale. In particular, $\bar Q_\text{Segmentation}$
decays exponentially with $n$ quite precisely, which is visible
in the logarithmic plot; this is exactly the behaviour predicted by
our simpler model.

Figure~\ref{vulnCompareFig} compares the data for $\bar Q_\text{Vulnerability}$.
For this parameter, we also obtain a good fit between the two models
across the range considered for $n$, except perhaps for the case of very few nodes ($n=50$).

\begin{figure} 
\begin{center}
\begin{minipage}{0.7\textwidth}
\resizebox{\textwidth}{!}{
\includegraphics*{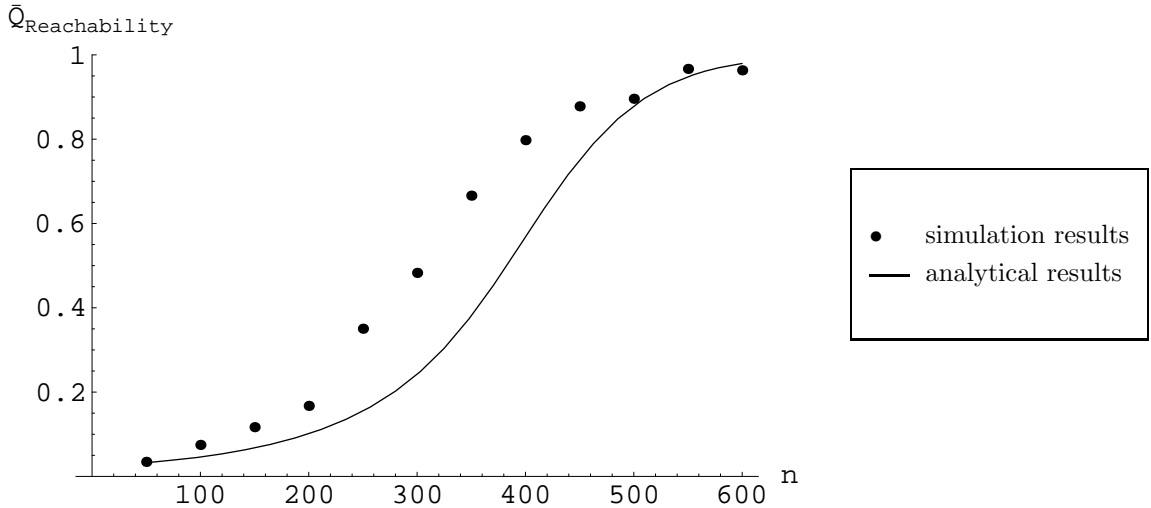}
}
\end{minipage}
\skywaysNumLegend
\end{center}
\caption{Analytical and simulation results for the reachability parameter}
\label{reachCompareFig}
\end{figure}

The last parameter -- reachability -- is shown in Fig.~\ref{reachCompareFig}.
While the qualitative behaviour agrees between the models also in this case,
there are noticeable differences in the absolute value of $\bar Q_\text{Reachability}$:
In the range of medium $n$, it seems that in the simple 1-dimensional model,
approximately 50-100 nodes more are needed to achieve the same reachability
as in the simulation by Roth. This leads to absolute differences of up to 
$0.3$ in $\bar Q_\text{Reachability}$ between the models.
Taking into account that the average \emph{number} of network segments
agrees between the models (cf.~Fig.~\ref{segmentationCompareFig}), 
this points to the fact that at least some particularly large segments
occurred in Roth's simulation that are not predicted by our 1-dimensional
model. This is possibly explained by the fact that Roth's model allows
communication between parallel paths; while we compensated this partially
by counting only one contribution to $\ell$ from two parallel paths, 
this still amounts to an increased density of nodes in those areas
that would not correctly be described by an equal distribution.

Certainly, it would be possible to gain a better and more quantitative
understanding of the difference between the two models by repeating and
modifying the simulations of \cite{Rot:critical_mass}, and by 
refining the construction in Chap.~\ref{ConnectChap} and~\ref{QualityChap}
in order to include more complex situations. However, such an analysis lies beyond
the scope of the current work.

In conclusion, it seems that the numerical results in \cite{Rot:critical_mass}
can be reproduced in our more simple model at least in a 
qualitative sense, and in large parts also quantitatively. 
It should be emphasized that this does not
amount to a comparison with experiment; we merely compared our results
to a different mathematical model, which is partially based on the
same simplifying assumptions (e.g. a homogeneous radio range for all nodes).
Still, the material of this section may support the claim that the predictions
of our 1-dimensional system are stable with respect to some changes in 
the modelling decisions. Differences with respect to details of the propagation model
could be compensated by a simple change in the system parameters.

\section{Quantitative predictions} \label{quantitSec}

More explicitely than the results known in the literature,
our asymptotic approximations allow us to make quantitative predictions
for the quality of 1-dimensional MANETs under the given modelling
assumptions, or, more importantly, to find appropriate system parameters
required to reach a certain quality level. This section gives some
examples to that end.

Assume in the following that the length $\ell$ of the MANET and the radio
range $r$ are given. We want to find the minimum node number $n$
needed to obtain different quality levels, 
where we restrict our attention to the case of large MANETs;
i.e. we will use the asymptotic formulas
for quality parameters from Table~\ref{qResultTable} on page~\pageref{qResultTable}. 

Let us start with connectedness. Given some required quality level $\bar Q_\text{Connectedness}$,
we can directly obtain the associated value $\eta$ by $\eta = -\ln(-\ln \bar Q_\text{Connectedness})$.
It remains to find $n$ such that $\eta = n r/\ell - \ln n$.
Given $r/\ell$, this solution needs to be calculated numerically, 
which is however easy to do (e.g. using Newton's algorithm).

For area coverage and segmentation, the required value of $\nu$ and hence of $n = \nu/r$
is directly obtained from $\bar Q_\text{Coverage}$ and $\bar Q_\text{Segmentation}$
without further complications. For the vulnerability, we need a numerical inversion
of $\nu \mapsto (\nu-1)e^{-\nu} + e^{-2\nu}$ in order to 
obtain $n$ from $\bar Q_\text{Vulnerability}$. 
(One usually obtains two such solutions for $n$ -- cf. Fig.~\ref{vulnCompareFig} --,
where we are interested in the greater one.)
Likewise, for the reachability parameter,
a numerical inversion of $x \mapsto 2x-(1+2x) e^{-1/x}$ gives us the
required value of $x = e^\eta$; we then proceed as above in order to
calculate $n$ from $\eta$.

\begin{table}
\centering
\setlength{\extrarowheight}{2pt}
\begin{tabular}{ p{0.3\textwidth}|r|r|}
criterion
&
\multicolumn{2}{|c|}{minimal node number $n$}
\\
&
\parbox{0.25\textwidth}{
\PBS\centering 
IEEE 802.11 WLAN

($r=\unit{30}{m}$)
}
&
\parbox{0.25\textwidth}{
\PBS\centering 
Bluetooth 

($r=\unit{10}{m}$)
}
\\
\hline
\hline
$\bar Q_\text{Connectedness} \geq 0.9$
&
261
&
906
\\
\hline
$\bar Q_\text{Coverage} \geq 0.9$
&
39
&
116
\\
\hline
$\bar Q_\text{Segmentation} \leq 0.1$
&
77
&
231
\\
\hline
$\bar Q_\text{Vulnerability} \leq 0.1$
&
102
&
304
\\
\hline
$\bar Q_\text{Reachability} \geq 0.9$
&
173
&
650
\\
\hline
\hline
$\bar Q_\text{Connectedness} \geq 0.99$
&
349
&
1167
\\
\hline
$\bar Q_\text{Coverage} \geq 0.99$
&
77
&
231
\\
\hline
$\bar Q_\text{Segmentation} \leq 0.01$
&
154
&
461
\\
\hline
$\bar Q_\text{Vulnerability} \leq 0.01$
&
209
&
627
\\
\hline
$\bar Q_\text{Reachability} \geq 0.99$
&
226
&
804
\\
\hline
\end{tabular}
\caption[Quantitative predictions for a 1-dimensional MANET]
{Quantitative predictions for a 1-dimensional MANET ($\ell = \unit{1000}{m}$).}
\label{quantitativeTable}
\end{table}

All these calculations can be performed with standard techniques
(Newton's method, \emph{regula falsi}) 
and without excessive need for computing capacity. In fact, 
the evaluation would be feasible even on a mobile device with 
very limited CPU power, should this become necessary e.g. within a 
distributed algorithm.

Table~\ref{quantitativeTable} shows some numerical examples for
a MANET of $\ell = \unit{1}{km}$ in length, using two different radio
ranges (for IEEE 802.11 WLAN and Bluetooth radios) and various quality criteria.
As expected, the non-intesive parameters (connectedness and reachability) lead to
criteria that are particularly demanding in terms of node density.
For example, if one requires 99\% probability of connectedness
in a Bluetooth-based MANET, then more than 1.100 network nodes are needed, 
which is more than one node per meter of network length -- a threshold
that would probably be hard to reach in practice.

\chapswitch
\chapter{Further Directions} \label{MiscChap}

This chapter discusses extensions of our results
to more complex situations. To that end, Sec.~\ref{VaryingNodeNumberSec}
presents a variation of our 1-dimensional MANET model in which the
network nodes may be switched off at random. 
Sec.~\ref{outlookSec} then gives a summary of the results
obtained in the current work, as well as an outlook 
to higher-dimensional systems and the description
of time dependence.

\newcommand{\vn}{\mathrm{VN}}

\section{A network with varying node number} \label{VaryingNodeNumberSec}

As a simple example of how our method can be generalized to more complex
behaviour, let us consider the following situation: In the 1-dimensional MANET,
we introduce a varying node number
by allowing each network node to be switched off at random. 
This corresponds to a user turning off their device e.g. for power saving
reasons. We will assume that at any fixed time, each device is 
switched on with probability $p$ (where the devices are independent of each other).
This is reflected in the model by adding a sample space 
$\Omega_\text{internal} = \{0,1\}$ for each node, where the value $0$ 
corresponds to the device being switched off. We thus consider 
the sample space
\begin{equation}
   \Omega_{n,\vn} = \big( [0,1] \times \{0,1\} \big)^n.
\end{equation}
We extend the probability measure by adding a discrete distribution
for each of the additional coordinates
$z_i \in \OmegaInt$ ($i = 1,\ldots,n$); the expectation value of a 
random variable $F_\vn : \Omega_{n,\vn} \to \rbb$
then is
\begin{equation} \label{vnExpect}
  \expect{F_\vn} = \sum_{z_1,\ldots,z_n=0}^1 \Big(\prod_{i=1}^n p^{z_i} (1-p)^{1-z_i}  \Big)
    \intcube d^nx\; F_\vn(x_1,\ldots,x_n,z_1,\ldots,z_n).
\end{equation}
Following our motivation, we can define our random variables of interest
(i.e. the quality parameters) quite easily: We want that for our quality measures,
only those nodes with $z_i=1$ are counted. That is, for a given family
of random variables $F^{(n)}: \Omega_n \to \rbb$ on the original MANET 
(with fixed node number), we define a variable $F_\vn$ on the new sample
space $\Omega_{n,\vn}$ by
\begin{equation}
  F_{\vn}^{(n)} (\xv,\zv) = F^{(n')} (\yv),
\end{equation}
where $n' = \sum_i z_i$, and $\yv = (y_1, \ldots, y_{n'})$ lists
those variables $x_i$ for which $z_i =1$. This definition is unambiguous if the
$F^{(n)}$ are symmetric, which was the case for all our quality parameters.

For this specific choice of random variable $F_\vn$, the expectation
value from Eq.~\eqref{vnExpect} is somewhat simplified: We can integrate
over all variables that do not appear in $F^{(n')}$, and make use of the fact that
$F^{(n')}$ does not depend on the $z_j$. This leads us to
\begin{equation}
  \expect{F_\vn} = \sum_{z_1,\ldots,z_n=0}^1 
  p^{\sum_{i=1}^{n} z_i}  (1-p)^{n-\sum_{i=1}^{n} z_i} 
    \int_{[0,1]^{n'}} d^{n'}y\; F^{(n')}(\yv).
\end{equation}
Since only the sum of the $z_i$ is relevant in this expression, we can 
replace the multiple sum by a single sum over $n'$:
\begin{equation} \label{transferExpect}
  \expect{F_\vn} = \sum_{n'=0}^n 
  \binom{n}{n'}
  p^{n'}  (1-p)^{n-n'} 
    \expect{F^{(n')}}.
\end{equation}

Clearly, one would expect that for large $n$, the MANET with varying node
number will behave like the MANET with fixed node number, but at the parameter
value $pn$ in place of $n$. 
Mathematically, this is a consequence of the central limit theorem.
We shall show this precisely at least for some parameters
of interest.

\begin{thm}
Let $Q^{(n,\rho)}$ be a family of random variables for the 1-dimensional MANET;
assume that $Q^{(n,\rho)}$ is scaling, symmetric, and intensive with limit
function $\tilde Q$. Moreover, let $Q^{(n,\rho)}$ be bounded in the sense that
there exists a constant $M>0$ such that
\begin{equation*}
   \forall n \in \nbb \; \forall \rho \in \rbb^+  \;\forall \omega \in \Omega_n:
   \;
   |  Q^{(n,\rho)}(\omega) | < M,
\end{equation*}
and suppose that the convergence $Q \to \tilde Q$ is uniform in the 
following sense:
\begin{equation*}
  \forall \epsilon > 0  \; \exists n_0  \; \exists \delta > 0  \;
  \forall n \geq n_0  :
  \quad
  |n \rho - \nu| < \delta  \;\Rightarrow\;  | \expect{Q^{(n,\rho)}} - \tilde Q(\nu) | < \epsilon. 
\end{equation*}
Let $Q_\vn^{(n,\rho)}$ be the corresponding random variable for the MANET
with varying node number. 
Then, for each sequence $(\rho_n)$ with $n \rho_n \to \nu > 0$, one has
\begin{equation*}
    \expect{Q_\vn^{(n,\rho_n)}} \to \tilde Q (p\nu).
\end{equation*}
\end{thm}

Note: It can easily be shown that the above condition of uniformity 
is fulfilled in all our examples of intensive quality parameters. 
Also, all our parameters were bounded by definition. 
An analogous
theorem for our non-intensive parameters might be stated, but we will not discuss
this in detail.

\begin{proof}
In view of Eq.~\eqref{transferExpect}, our task is to show that
\begin{equation}
  \sum_{n'=0}^n 
  \binom{n}{n'}
  p^{n'}  (1-p)^{n-n'} 
    \big| \expect{ Q^{(n',\rho_n)}}  - \tilde Q (p\nu)  \big|  \longnlim 0.
\end{equation}
To that end, let $\epsilon >0$ be given. Further, let $\lambda > 0$
(its value will be specified later). Let $\sigma_n = \sqrt{np(1-p)}$,
$\alpha_n = [np - \lambda \sigma_n]$, and $\beta_n = [np + \lambda \sigma_n]$.
Applying the de-Moivre-Laplace theorem (cf. Theorem~\ref{dmlThm} in Appendix~\ref{LimitApp}),
we know that
\begin{equation}
 \lim_{n \to \infty} \sum_{n'=0}^{\alpha_n} \binom{n}{n'} p^{n'} (1-p)^{n-n'} = \Phi(-\lambda).
\end{equation}
(See Eq.~\eqref{phiDef} for the definition of $\Phi$.) Since with $Q$, also its
limit function $\tilde Q$ must be bounded, we can thus obtain for large $n$:
\begin{equation} \label{nprimeLowerSum}
  \sum_{n'=0}^{\alpha_n}
  \binom{n}{n'}
   p^{n'}  (1-p)^{n-n'} 
    \big| \expect{ Q^{(n',\rho_n)}}  - \tilde Q (p\nu)  \big|  \leq 2 M (\Phi(-\lambda) + \epsilon/M).
\end{equation}
Likewise, we see that
\begin{equation} \label{nprimeUpperSum}
  \sum_{n'=\beta_n+1}^{n}
  \binom{n}{n'}
   p^{n'}  (1-p)^{n-n'} 
    \big| \expect{ Q^{(n',\rho_n)}}  - \tilde Q (p\nu)  \big|  \leq 2 M (\Phi(-\lambda) + \epsilon/M).
\end{equation}
It remains to control the sum over $n' \in \{\alpha_n+1,\ldots,\beta_n\}$.
For these values of $n'$, we certainly have
\begin{equation}
 | n' \rho_n  - p \nu | \leq \rho_n | n' - np| + p | n \rho_n - \nu|
 \leq \lambda \rho_n \sigma_n + p |n \rho_n - \nu |.
\end{equation}
Since $\rho_n = \Theta(1/n)$,  $\sigma_n = \Theta(\sqrt{n})$, and $n \rho_n \to \nu$,
we can achieve that  $ | n' \rho_n  - p \nu | < \delta$ for sufficiently large $n$,
where $\delta$ is the value used in the uniformity assumption. This assumption 
then guarantees that $|\expect{ Q^{(n',\rho_n)}}  - \tilde Q (p\nu)| < \epsilon$
for large $n$ and $\alpha_n < n' \leq \beta_n$; thus
\begin{equation} \label{nprimeMidSum}
  \sum_{n'=\alpha_n+1}^{\beta_n}
  \binom{n}{n'}
   p^{n'}  (1-p)^{n-n'} 
    \big| \expect{ Q^{(n',\rho_n)}}  - \tilde Q (p\nu)  \big|  \leq \epsilon.
\end{equation}
Combining Eqs.~\eqref{nprimeLowerSum}, \eqref{nprimeUpperSum}, and \eqref{nprimeMidSum},
and choosing $\lambda$ large enough such that $\Phi(-\lambda) < \epsilon/M$, we have
achieved the desired result.
\end{proof}

The above theorem says that for the model with varying node number,
we can apply the results from Chap.~\ref{QualityChap} directly if we set 
the node number in those results to $np$ (i.e. to its statistical mean).
While this is not very surprising, it means that our model is stable (to some extent)
against changes in the assumptions; we can accommodate the extra effect
by merely modifying one of the system's parameters.

For illustration, let us discuss the above findings in one concrete example,
namely the segmentation parameter $Q_\text{Segmentation}$ introduced in 
Sec.~\ref{SegmentationSec}. Here we know from Eq.~\eqref{segmentResult} that
\begin{equation}
  \expect{Q_\text{Segmentation}} = (1-\rho)^{n-1}.
\end{equation}
Inserting into Eq.~\eqref{transferExpect}, we can explicitly calculate the segmentation
for varying node number:
\begin{multline}
  \expect{Q_{\text{Segmentation},\vn}} = \sum_{n'=0}^n 
  \binom{n}{n'}
  p^{n'}  (1-p)^{n-n'} (1-\rho)^{n'-1}
  \\
 = \frac{1}{1-\rho} \big( p(1-\rho) + (1-p) \big) ^n
 = \frac{(1-p\rho)^n}{1-\rho}
\end{multline}
In the limit $n\rho \to \nu$, it follows that
\begin{equation}
  \expect{Q_{\text{Segmentation},\vn}} 
   \to e^{-p\nu}  = \tilde Q_\text{Segmentation} (p\nu),
\end{equation}
as expected.

\section{Conclusions and outlook} \label{outlookSec}

In the course of the present work, we have analysed a 1-dimensional
MANET system with statistical methods. Using a number of symmetries
of the system, the mathematical description of connectivity
properties could be much simplified. It turned out that 
the model was explicitly solvable when boundary effects were neglected
(through the use of periodic boundary conditions). 
In particular, we were able to obtain an explicit expression for
the probability of connectedness for given parameters, and analyse
this expression in the limit of large MANET size. This improves
the results known in the literature for 1-dimensional systems.

We then analysed a number of different quality measures for MANETs. 
In general, quality parameters could be classified into intensive 
parameters (with good scaling properties)
and non-intensive ones (which possibly lead to scalability problems).
We were able to obtain explicit results for all of the parameters
in the simple 1-dimensional model. Our results agree with the numerical
data known in the literature.

Our results can serve both as a qualitative and quantitative guideline
for the design of 1-dimensional MANET systems, in particular for sensor networks.
Due to our explicit results for the expectation
value of quality parameters, it is easy to choose the radio range or
node density in a MANET such that it reaches the desired quality level.
In particular, this applies to the asymptotic formulas; 
they are certainly simple enough to even allow computation on the mobile
devices themselves.

Further, the methods we have developed should be applicable also to
other quality parameters, in case they are desired for specific applications:
As long as these parameters can reasonably be expressed in terms of
the next-neighbour coordinates, it should be possible to apply the 
techniques of Chap.~\ref{QualityChap} in order to obtain their expectation value.

Certainly, we have merely treated a small part of the problems and 
obstacles that may limit the quality and scalability of MANETs. 
In particular, we have not dealt with questions of routing, throughput,
and all aspects explicitly related to mobility. Thus, our results
should be regarded as a \emph{upper bound} to MANET quality, in the sense
that additional problems might be faced on higher layers.

Our specific 1-dimensional model is quite simplistic in its assumptions,
and it would certainly be worthwhile to study some extensions in order to 
explore the stability of our results against changes in the model. 
Apart from an inhomogeneous spatial distribution of the nodes, 
it would be particularly interesting to analyse nodes with a varying
radio range, which might be caused e.g. by local interference, changes in antenna
positions, or shielding. In analogy to Sec.~\ref{VaryingNodeNumberSec},
this could be modelled by introducing additional random parameters into the formalism
which control e.g. the radio range between each pair of nodes, 
or only between next neighbours. Still, one would expect that under reasonable
assumptions, the extended model could effectively been reduced to the 
known situation by application of the central limit theorem.

It would also be desirable to extend our findings to 2-dimensional and,
with some limitations, to 3-dimensional MANET systems. In fact,
some of the results can easily be generalized: Let us consider the area coverage
parameter $Q_\text{Coverage}$. Assume that $n$ nodes with circular radio range $r$ 
are distributed equally (and independently) to the cube $[0,\ell]^d$
($d \in \nbb$), considered with periodic boundary conditions.
We can certainly say that $\expect{1-Q_\text{Coverage}}$
is the probability that a dedicated point, distributed at random to 
$[0,\ell]^d$, will fall into the range of \emph{none} of the MANET nodes.
The probability for the dedicated point to fall into the range of one specific
node, however, is simply $c_d \rho^d$, where $\rho=r/\ell$ as usual, and $c_d$ is the
volume of the unit sphere in $d$ dimensions. Due to the independent distribution
of network nodes, we obtain
\begin{equation}
  \expect{Q_\text{Coverage}} = 1 - (1- c_d \rho^d)^n \to 1 - e^{-c_d\nu}
  \quad
  \text{ as } n \rho^d \to \nu,
\end{equation}
in generalization of our 1-dimensional result in Eq.~\eqref{coverageResult1d};
we have
\begin{equation}
  c_1 = 2, \quad
  c_2 = \pi, \quad
  c_3 = \frac{4}{3} \pi.
\end{equation}

The results for other quality measures, in particular for connectedness,
do not transfer that obviously however: Since the next-neighbour coordinates
cannot be used in the same way in higher dimensions, we would first have
to find appropriate new coordinates in order to transfer our methods. 
On the other hand, similar results would be expected to hold;
cf. the numerical results by Santi and Blough~\cite{SanBlo:transmitting_range}
and the analytical estimates by Bettstetter~\cite{Bet:minimum_node_degree}.

\begin{figure} 
\centering
  \resizebox{!}{0.3\textheight}{
      \includegraphics{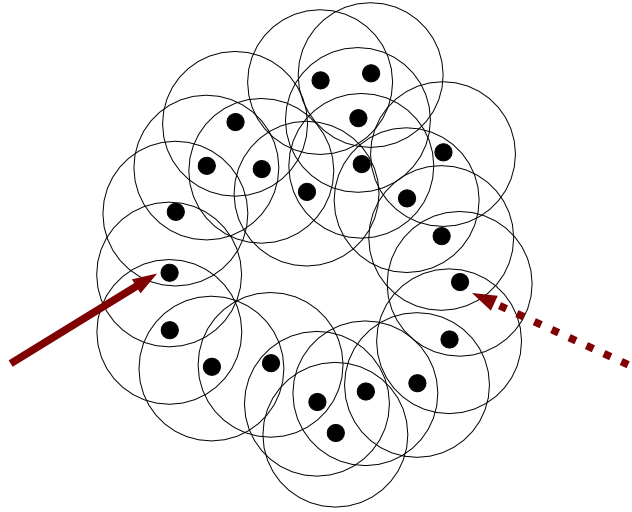}
  }
\caption[Long-range dependence of the vulnerability in 2 dimensions]{
Long-range dependence of the vulnerability in 2-dimensional networks. The importance
of the node marked with a solid arrow depends on the position of the node
marked with a dotted arrow, and vice versa.} 
\label{importanceFig}
\end{figure}

It should also
be noted that certain quality measures somewhat change their nature in $d>1$
dimensions: As an example, consider the vulnerability parameter (cf. Sec.~\ref{VulnSec}).
In the 1-dimensional situation, the question whether a node is ``important'' 
for network connectivity is determined by its two associated next-neighbour
distances, and hence we may say that it is a local property. In $d \geq 2$, however, it may
happen that the importance of a node depends on the structure of the network
at a very remote place (see Fig.~\ref{importanceFig}). 
Since we want our quality measures to reflect the behaviour in the bulk network,
it might even be necessary to change the definition of the quality parameters
in higher dimensions.

Up to now, we have only considered static deployments of network nodes,
taking account for mobility only through our assumption of ergodicity.
While for the quality parameters we considered, we are in good agreement
with simulations that rely on an explicit motion model (cf. Sec.~\ref{SkywaysSec}),
our framework is certainly not useful for determining quality measures that are
directly linked to the time evolution of the system, such as the question:
``What is the probability that the network is connected for a time frame
of length $t_0$?'' In order to answer such questions, we need to 
make specific assumptions on the motion of nodes. 

Certainly, it would be possible to incorporate one of the common explicit
mobility models, like random waypoint or Brownian motion, into our context.
From a more general point of view, however, these explicit models seem to be
rather ad hoc and include some aspects that are not really motivated by
properties of the real network system (such as discrete time steps).
These technicalities could even complicate an explicit analysis more
than necessary.
Therefore, it might be desirable to consider a model with more natural
assumptions, or explore and compare several such modelling alternatives.

On the mathematical side, passing to such an analysis -- without assuming
a discrete time scale -- would mean that we pass from our finite-dimensional
sample space $\Omega_n$ to an infinite-dimensional space of functions.
The base for such an \emph{ab origine} calculation could be found in the 
theory of stochastic integrals and stochastic differential equations;
while this field is well established \cite{Pro:stochastic, PKL:stochastic},
it would certainly increase the technical complexity of our analysis by far,
compared with the rather elementary mathematical methods used in the present
work. Still, this might be a promising subject for future research.

\chapswitch
\appendix
\renewcommand{\chaptername}{Appendix}

\chapter{Some Mathematical Machinery} \label{MathApp}

\section{The standard simplex in higher dimensions} \label{SimplexApp}

In our analysis, we often deal with a specific volume in $\rbb^n$,
the \emph{$n$-dimensional standard simplex,} defined as
\begin{equation} \label{simplexDef}
  V_n := \big\{ \xv \in [0,1]^n \,\big |\, \sum_{i=1}^n x_i \leq 1  \big\}.
\end{equation}
We are also often lead to the \emph{top surface} of $V_n$, which we 
denote by $T_n$ and define it as
\begin{equation}\label{simplexSurfaceDef}
  T_n := \big\{ \xv \in [0,1]^n \,\big |\, \sum_{i=1}^n x_i = 1  \big\}.
\end{equation}
$T_n$ is an $(n-1)$-dimensional manifold spanned by $n$ corner points, which are 
all located at equal mutual distances; specifically, 
$T_1$ is a single point, $T_2$ a straight line,
$T_3$ an equilateral triangle, and $T_4$ a tetrahedron.

In this appendix, we discuss several properties of $V_n$ and $T_n$, where it seems
appropriate to develop them separately from the main text.
Specifically, we calculate certain integrals over $V_n$ and $T_n$ that turn out
to be important for our argumentation.

First of all, let us introduce Heaviside's theta function
\begin{equation} \label{thetaDef}
  \theta(x) := \begin{cases}
     1 & \text{if } x \geq 0, \\
     0 & \text{if } x < 0;
    \end{cases}
\end{equation}
note that $\theta(x) = \theta(\lambda x)$ for all $\lambda \in \rbb^+$, $x \in \rbb$.
Also, we have $\theta(-x) = 1-\theta(x)$ except for $x=0$ (this set of zero volume
can be neglected in integrals).
Using the $\theta$ function, we can express the integral of any function $f$ 
over $V_n$ as follows:
\begin{equation}\label{vnint}
   \int_{V_n} d^nx\, f(x) =  \intcube d^nx \, f(x) \, \theta\big( 1 - \sum_{i=1}^n x_i \big).
\end{equation}
We will now calculate certain integrals over $V_n$ explicitly.
\begin{lemm} \label{vnIntLemm}
For any $n \in \nbb$ and $k \in \nbb_0$, we have
\begin{equation*}
 \int_{V_n} d^nx \; \big(1-\sum_{i=1}^n x_i \big)^k = \frac{k!}{(k+n)!}.
\end{equation*}
\end{lemm}

\begin{proof}
We will prove the relation by induction on $n$. For $n=1$, the proposition reads
\begin{equation}
 \int_{0}^{1} dx \; (1-x)^k = \frac{1}{k+1},
\end{equation}
which is easily checked by direct calculation. Now let the proposition
be true for $n-1$ in place of $n$, with $k \in \nbb_0$ being arbitrary.
Using Eq.~\eqref{vnint}, we calculate
\begin{multline} \label{vnbasecalc}
 \int_{V_n} d^nx \; \big(1-\sum_{i=1}^n x_i \big)^k
 =  \int_{ [0,1]^{n-1}} d^{n-1}x \int_0^1 dx_n  \big(1-\sum_{i=1}^n x_i \big)^k
   \theta\big( 1-\sum_{i=1}^{n-1} x_i - x_n \big)
 \\
 =  \int_{ [0,1]^{n-1}} d^{n-1}x \;
 \theta\big( 1-\sum_{i=1}^{n-1} x_i \big)
  \int_0^{1-\sum_{i=1}^{n-1} x_i} dx_n \; \big(1-\sum_{i=1}^n x_i \big)^k.
\end{multline}
Setting $a = 1-\sum_{i=1}^{n-1} x_i$, the integral in $x_n$ can be elementary solved as
\begin{equation}
\int_0^{a} dx_n  \big(a-x_n)^k
 = \frac{1}{k+1} \big\lbrack -(a-x_n)^{k+1} \big\rbrack_{0}^{a}
= \frac{1}{k+1} a^{k+1}.
\end{equation}
Inserting this result into Eq.~\eqref{vnbasecalc}, we have
\begin{equation}
 \int_{V_n} d^nx \; \big(1-\sum_{i=1}^n x_i \big)^k
  = \int_{V_{n-1}} d^{n-1}x  \frac{1}{k+1} \big(1-\sum_{i=1}^{n-1} x_i \big)^{k+1}.
\end{equation}
By induction hypothesis, this evaluates to
\begin{equation}
 \int_{V_n} d^nx \; \big(1-\sum_{i=1}^n x_i \big)^k
  = \frac{1}{k+1}  \frac{(k+1)!}{(n-1+k+1)!} = \frac{k!}{(n+k)!},
\end{equation}
as desired.
\end{proof}

Our next task is to calculate similar integrals over the top surface $T_n$ of
$V_n$. For calculating such an integral of some function $f$,
\begin{equation}
    \int_{T_n} dS(\xv)  f(\xv),
\end{equation} 
where $dS(\xv)$ is the surface element of $T_n$, we need a coordinatization 
of the surface and the length of its normal vector. Since the surface
is characterized by the equation
\begin{equation}
 1 - \sum_{i=1}^n x_i = 0,
\end{equation}
coordinates are simply given
by e.g. $(x_1,\ldots,x_{n-1}) \in V_{n-1}$, 
setting $x_n = 1 - \sum_{i=1}^{n-1} x_i$, 
and the normal vector is easily seen to be $(1,1,\ldots,1) \in \rbb^n$,
so its length is $\sqrt{n}$. Thus, 
for $n \geq 2$, we can calculate the integral as
\begin{equation} \label{surfaceIntNoDelta}
    \int_{T_n} dS(\xv) \; f(\xv) 
    = \sqrt{n} \int_{V_{n-1}} d^{n-1}x \; f(x_1,\ldots,x_{n-1},1 - \sum_{i=1}^{n-1} x_i).
\end{equation} 
For $n=1$, the surface $T_n$ is a single point, and we have $\int_{T_1} dS(\xv) f(\xv) = f(1)$.
We shall often represent the integral in a different way:
Using the ``delta valued measure'' concentrated on $T_n$, we can rewrite Eq.~\eqref{surfaceIntNoDelta} as
\begin{equation}\label{surfaceIntDelta}
    \int_{T_n} dS(\xv)  f(\xv) = \sqrt{n} \intcube d^nx \, \delta(1-\sum_{i=1}^{n} x_i) f(\xv).
\end{equation}
which also holds for $n=1$. In many situations, we prefer the latter form of notation, since it
expresses the symmetry between the $n$ different coordinates more directly.
The reader unfamiliar with delta-valued measures \cite{GS:distributionsOneTwo} can always replace this
expression with \eqref{surfaceIntNoDelta} if in doubt.

We will now calculate some commonly used surface integrals.

\begin{lemm} \label{tnIntLemm}
For any $n \in \nbb$, $k \in \nbb_0$, and $j \in \{1,\ldots,n\}$, we have
\begin{equation*}
  \int_{T_n} dS(\xv) x_j^{k} = \frac{ \sqrt{n} \, k!}{(n+k-1)!}.
\end{equation*}
\end{lemm}
\begin{proof}
The statement is easily checked for $n=1$; so let $n \geq 2$ in the following.
Due to symmetry reasons, we can choose $j=n$ without loss of generality.
Now setting $f(\xv)=x_n^k$ in Eq.~\eqref{surfaceIntNoDelta}, we see that
\begin{equation}
  \int_{T_n} dS(\xv) x_n^{k} = \sqrt{n} \int_{V_{n-1}} d^{n-1}x \; (1 - \sum_{i=1}^{n-1} x_i)^k.
\end{equation}
The integral on the right-hand side is known by Lemma~\ref{vnIntLemm}; inserting 
that expression, we can immediately show the proposed result.
\end{proof}

Let us note some consequences of the previous lemmas: Setting $k=0$
in Lemma~\ref{vnIntLemm}, we can calculate the volume of $V_n$ as
\begin{equation} \label{volvn}
	\vol(V_n) = \frac{1}{n!}.
\end{equation}
In the same way, setting $k=0$ in Lemma~\ref{tnIntLemm}, we can
determine the $(n-1)$-dimensional volume of $T_n$ as
\begin{equation} \label{voltn}
	\vol(T_n) = \frac{\sqrt{n}}{(n-1)!}.
\end{equation}
By the latter result, we can easily write down the probability measure
of equal distribution on the surface $T_n$, which fulfills 
$d\muEqualT{n}(\xv) = (\vol T_n)^{-1} dS(\xv)$. We can summarize this
as follows.
\begin{prop} \label{tnEqualProp}
The measure of equal distribution over the surface $T_n$ has the form
\begin{equation}
\muEqualT{n}(\xv) = (n-1)! \, \delta(1-\sum_{i=1}^{n} x_i),
\end{equation}
considered on the space $[0,1]^n$. 
For any $n \in \nbb$, $k \in \nbb_0$, and $j \in \{1,\ldots,n\}$, we have
\begin{equation*}
  \intcube d\muEqualT{n}(\xv) \; x_j^{k} = \binom{k+n-1}{k}^{-1}.
\end{equation*}
\end{prop}
The second part of the proposition follows directly from Lemma~\ref{tnIntLemm}
and Eq.~\eqref{voltn}. We now turn to another often-used relation,
which might be described as a scaling argument on $T_n$. To that end,
let $\unitv{j} = (0,\ldots,0,1,0,\ldots,0)$ (with the $1$ in the $j$-th place)
denote the $j$-th standard unit vector in $\rbb^n$. 

\begin{lemm} \label{tnScalingLemm}
Let $n \in \nbb$, $j \in \{1,\ldots,n\}$, and $\lambda \in (0,1)$,
 and let $f : T_n \to \rbb$
be integrable. Then
\begin{equation*}
\intcube d\muEqualT{n}(\xv) \; \theta(x_j-\lambda) f(\xv)
= (1-\lambda)^{n-1} 
\intcube d\muEqualT{n}(\xv) \;
f( (1-\lambda)\xv + \lambda \unitv{j}).
\end{equation*}
\end{lemm}

\begin{proof}
Since the integration measure does not change when permuting the variables,
we can assume without loss of generality that $j=n$. 
By Eq.~\eqref{surfaceIntNoDelta}, we have
\begin{multline}
\intcube d\muEqualT{n}(\xv) \; \theta(x_n-\lambda) f(\xv)
\\
= (n-1)! \int_{V_{n-1}} d^{n-1}x \;
    \theta(1-\sum_{i=1}^{n-1}x_i-\lambda) \;
    f(x_1,\ldots,x_{n-1}, 1-\sum_{i=1}^{n-1}x_i)
\\
= (n-1)! \intcube d^{n-1}x \;
    \theta(1-\sum_{i=1}^{n-1}x_i) \;
    \theta((1-\lambda)-\sum_{i=1}^{n-1}x_i) \;
    f(x_1,\ldots,x_{n-1}, 1-\sum_{i=1}^{n-1}x_i).
\end{multline}
Since $(1-\lambda) < 1$, the first theta function is redundant
in view of the second one. Then, a variable transformation
$x_i' = (1-\lambda)^{-1} x_i$ leads us to
\begin{multline}
\intcube d\muEqualT{n} \; \theta(x_n-\lambda) f(\xv)
\\
= (n-1)! \;(1-\lambda)^{n-1} \int_{V_{n-1}} d^{n-1}x' \;
    f((1-\lambda)x_1',\ldots,(1-\lambda)x_{n-1}', 1-\sum_{i=1}^{n-1}(1-\lambda)x_i')
\\
= (1-\lambda)^{n-1} \intcube d\muEqualT{n} (\xv' )\;
    f( (1-\lambda)\xv' + \lambda \unitv{n}),
\end{multline}
which was to be shown.
\end{proof}

Using the previous lemma, we will establish a related
technical result which turns out to be useful for our purposes.

\begin{lemm} \label{tnDoubleThetaLemm}
Let $n \in \nbb$, $n \geq 2$, and let 
$j,k \in \{1,\ldots,n\}$ with $j \neq k$;
furthermore, let $\lambda \in (0,1)$.
Then
\begin{equation*}
\intcube d\muEqualT{n} \; \theta(\lambda-x_j-x_k) 
= 1-(1-\lambda)^{n-2} (1+ (n-2)\lambda).
\end{equation*}
\end{lemm}

\begin{proof}
In the case $n =2$, both sides of the proposed relation vanish;
so let $n \geq 3$.
Without loss of generality, we may assume $j=n-1$ and $k=n$.
Observe that
\begin{multline}
\intcube d\muEqualT{n}(\xv) \; \theta(\lambda-x_{n-1}-x_n) 
\\
= (n-1)! \int_{V_{n-1}} d^{n-1}x  \;
  \theta(\lambda-x_{n-1}-(1-\sum_{i=1}^{n-1}x_i)) 
\\
= (n-1)! \int_{V_{n-2}} d^{n-2}x  \;
  \theta(\lambda-1+\sum_{i=1}^{n-2}x_i) 
  \int_0^{1-\sum_{i=1}^{n-2}x_i} dx_{n-1}
\\
= (n-1)! \int_{V_{n-2}} d^{n-2}x  \;
  \theta(\lambda-(1-\sum_{i=1}^{n-2}x_i)) \;
  (1-\sum_{i=1}^{n-2}x_i).
\end{multline}
We can rewrite this expression as an integral over $T_{n-1}$:
\begin{multline}
\intcube d\muEqualT{n}(\xv) \; \theta(\lambda-x_{n-1}-x_n) 
%
%\\
= \frac{(n-1)!}{(n-2)!} \intcube  d\muEqualT{n-1}(\xv) \;
  \underbrace{\theta(\lambda-x_{n-1})}_{ 1 - \theta(x_{n-1}-\lambda)} \; 
  x_{n-1}
\\
= (n-1) \Big( \intcube  d\muEqualT{n-1}(\xv) \;
  x_{n-1}
-
   \intcube  d\muEqualT{n-1}(\xv) \;
  \theta(x_{n-1}-\lambda) \;
  x_{n-1} 
  \Big).
\end{multline}
The first integral expression is known by Proposition~\ref{tnEqualProp};
on the second one, we can apply Lemma~\ref{tnScalingLemm}. 
This yields
\begin{multline}
\intcube d\muEqualT{n}(\xv) \; \theta(\lambda-x_{n-1}-x_n) 
\\
= 1 - (n-1) (1-\lambda)^{n-2} \intcube  d\muEqualT{n-1}(\xv) \;
  (\lambda + (1-\lambda) x_{n-1}).
\end{multline}
Again applying Proposition~\ref{tnEqualProp}, our result is
\begin{multline}
\intcube d\muEqualT{n}(\xv) \; \theta(\lambda-x_{n-1}-x_n) 
= 1 - (n-1) (1-\lambda)^{n-2} (\lambda + \frac{1-\lambda}{n-1})
\\
= 1 - (1-\lambda)^{n-2} (1+(n-2)\lambda),
\end{multline}
as proposed.
\end{proof}

\section{The inclusion-exclusion formula} \label{IncExcApp}

At several points in the main text, we make use of the
well-known \emph{inclusion-exclusion formula,} which allows us
to calculate the probability of certain events easily.
We formulate it here for reference.

\begin{thm} \label{incExcThm}
Let $\Omega$ be a sample space, and let 
$A_1, \ldots, A_n \subset  \Omega$ be events in it.
For $k \in \nbb_0$, let
\begin{align*}
  B_k &:= \{ \omega \in \Omega \,|\, \omega \in A_j \text{ for exactly $k$ values of $j$} \},
  \; \text{ and }
  \\
  C_k &:= \{ \omega \in \Omega \,|\, \omega \in A_j \text{ for at least $k$ values of $j$} \}.
\end{align*}
Then we have
\begin{align*}
  P(B_k) &= \sum_{j=k}^n (-1)^{j-k} \binom{j}{k} S_j,
  \\
    P(C_k) &= \sum_{j=k}^n (-1)^{j-k} \binom{j-1}{k-1} S_j,
\end{align*}
where $S_j$ is defined as 
\begin{equation*}
	S_j := \sum_{ \{ m_1,\ldots,m_j \} } P(A_{m_1} \cap \ldots \cap A_{m_j});
\end{equation*}
the sum runs over all subsets $\{ m_1,\ldots,m_j \} \subset \{1,\ldots,n\}$.
\end{thm}
A proof of this formula can be found in most textbooks
on elementary statistics -- see, for example, 
the book by Krengel \cite[Sec.~3.4]{Kre:Statistik}.

\section{Statistical limits} \label{LimitApp}

In this appendix, we will state some familiar limit theorems
that are useful in our discussion; they are reproduced here
for easier reference. The first of these is Stirling's formula,
which gives an approximation of the factorial $n!$ for large $n$.
Its precise form is:
\begin{thm} \label{stirlingThm}
For each $n \in \nbb$, there is a $\delta(n) \in [\frac{1}{12n+1},\frac{1}{12n}]$
such that
\begin{equation*}
  n! = \sqrt{2 \pi n} \Big(\frac{n}{e}\Big)^n e^{\delta(n)}.
\end{equation*}
\end{thm}
A proof can be found e.g. in \cite[Appendix to \S 5]{Kre:Statistik}. It follows
in particular that $n! \geq (n/e)^n$ for all $n \in \nbb$; this is the inequality 
we will actually use.

The next result which we want to note (in fact a consequence of 
Theorem~\ref{stirlingThm})  is the Theorem of de Moivre-Laplace, which tells us about
the convergence of binomial probability distributions to normal (Gaussian)
distributions. To that end, let $F$ be a random variable which is 
binomially distributed with parameters $n$ and $p$; that is,
\begin{equation}
  \forall i \in \{0,\ldots,n\}:\;	P(F = i) = \binom{n}{i} p^i (1-p)^{1-i}.
\end{equation}
We write $\sigma_n = \sqrt{np(1-p)}$. Further, let the function $\Phi$ be defined as
\begin{equation} \label{phiDef}
  \Phi(x) = \frac{1}{\sqrt{2 \pi}} \int_{-\infty}^x dy \;e^{-y^2/2};
\end{equation}
we additionally set $\Phi(-\infty) = 0$ and $\Phi(+\infty) = 1$.
Note that $\Phi(-x) = 1 - \Phi(x)$ for all $x$. The de-Moivre-Laplace theorem
then states the following.

\begin{thm} \label{dmlThm}
   Let $F$ be as above, and let $\alpha \in \rbb \cup \{-\infty\}$, 
   $\beta \in \rbb \cup \{+\infty\}$, where $\alpha < \beta$. Then
   it holds that
\begin{equation*}
	\lim_{n \to \infty} P( np+ \alpha \sigma_n \leq F \leq np+ \beta \sigma_n )
	= \Phi(\beta) - \Phi(\alpha).
\end{equation*}   
\end{thm}

For a proof, again see \cite[\S 5]{Kre:Statistik}. It is well known that 
the above theorem is only a special case of the more general
\emph{central limit theorem;} however, we shall only need the specialized form
for our purposes.

\section{A summation lemma} \label{SummationApp}

This appendix presents an auxiliary result regarding a summation formula.
The idea is to express the sum in question as a 
power series of a certain function, then using well-known
relations for its derivatives in order to achieve the desired result.

\begin{lemm} \label{kDeltaSumLemm}
Let $j \in \nbb_0$. Then
\begin{equation*}
  \sum_{k=0}^j (-1)^k k \binom{j}{k} = 
  \begin{cases}
  	 -1 & \text{if } j=1, \\
  	 0 & \text{otherwise}.
  \end{cases}
\end{equation*}
\end{lemm}

\begin{proof}
 For $j=0$ and $j=1$, one checks by explicit calculation that the proposition is true.
 Now let $j \geq 2$, and let the function $f$ be defined as
 \begin{equation}
 	f(x) = (1-x)^j = \sum_{k=0}^j \binom{j}{k} (-x)^k.
 \end{equation}
 Then we know from the expression on the right-hand side that
 \begin{equation} \label{kSumDelta1}
 	\frac{df}{dx}\Big|_{x=1} = \sum_{k=0}^j (-1)^k \binom{j}{k} k x^{k-1}\Big|_{x=1}
 	= \sum_{k=0}^j (-1)^k k \binom{j}{k}.
 \end{equation}
On the other hand,
 \begin{equation}\label{kSumDelta2}
 	\frac{df}{dx}\Big|_{x=1} = -j(1-x)^{j-1} \Big|_{x=1}
= 0,
 \end{equation}
since $j>1$. Combining Eqs.~\eqref{kSumDelta1} and~\eqref{kSumDelta2},
we have proved the proposed result. 
\end{proof}

\chapswitch

\chapter{Notes on a Series of Publications by P.~Santi et al.} \label{SantiApp}

In a recent series of publications, 
P.~Santi and D.~M.~Blough
\cite{SanBlo:connectivity,SanBlo:transmitting_range},
as well as the same authors and F.~Vainstein \cite{SBV:range_assignment},
have analysed the connectedness problem of MANETs using statistical models.
Among others, they considered the very same mathematical model for
a 1-dimensional MANET (with disconnected boundary conditions) that we have
used in Sec.~\ref{disconnectedSec}. The authors proposed
a number of asymptotic estimates for the probability of connectedness;
however, as mentioned in Sec.~\ref{pconnLiteratureSec}, these estimates
are in the general case incompatible with the results of our analysis.
The present author claims that several theorems established in 
\cite{SBV:range_assignment,SanBlo:connectivity,SanBlo:transmitting_range} 
do in fact not hold in the form stated there; this appendix will discuss
counterexamples to those theorems, as well as pointing out inconsistencies 
in their corresponding proofs.

In the following, we shall stick to the notation used in \cite{SanBlo:transmitting_range}
rather than that used in the main text. This means in particular that
we regard $r$ and $\ell$ as two explicit parameters (rather than using
the normalized radio range), that we will consider $r$ and $n$ as functions of $\ell$,
and describe the limit of large systems as $\ell \to \infty$. (This
is only a question of nomenclature.)

Let us start with the upper bounds on the probability of connectedness
as proposed in \cite[Theorem~4]{SanBlo:transmitting_range}.
The authors state the following.

\begin{quotepar}
  ``Assume that $n$ nodes, each with transmitting range $r$, are distributed
  uniformly and independently at random in $R=[0,\ell]$ and assume that
  $rn=k \ell \ln \ell$ for some constant $k>0$. Further, assume that
  $r = r(\ell) \ll \ell$ and $n = n (\ell) \gg 1$. If $k > 2$,
  or $k=2$ and $r = r(\ell) \gg 1$, 
  then $\lim_{\ell \to \infty} P(CONN_\ell) = 1$.''
\end{quotepar}

(Here $\event{$\text{CONN}_\ell$}$ is the event $\conndb$ in our notation, and 
$r \ll \ell$ means $r/\ell \to 0$, etc.)
This statement is in conflict with our results: As a counterexample,
consider $r = \ell^{-k} \ln \ell$, $n = k \ell^{k+1}$, where $ k > 2$.
Then all conditions of the above statement are fulfilled; however, since
$\ln n \geq (k+1) \ln \ell$, one has
\begin{equation}
   n \;r \leq \frac{k}{k+1} \;\ell \;\ln n,
\end{equation}
and thus $P_\conndb \to 0$ according to Corollary~\ref{SantiCompareCorr}.

In fact, the proof of Theorem~4 in the Appendix of \cite{SanBlo:transmitting_range} 
is inconclusive: After Eq.~(2), the authors calculate the intermediate result
\begin{equation} \label{logESanti}
  \ln E[ \mu(n,C) ] < \ln \frac{2 \ell}{r} - \frac{k \ln \ell}{2}
   = \ln \frac{2}{r \ell^{k/2-1}},
\end{equation}
where $C = 2 \ell / r$, and $\mu(n,C)$ is a random variable whose details 
are not relevant here. Then they state:

\begin{quotepar}
  ``If $k > 2$, or if $k=2$ and $r = r(\ell) \gg 1$, then it is easily
  seen from this expression that 
  $\lim_{n,C \to\infty} \ln E[\mu(n,C)] = -\infty$.''
\end{quotepar}

However, this conclusion is not justified: In the above counterexample,
one has 
\begin{equation}
 \frac{2}{r \ell^{k/2-1}} = \frac{ 2 \;\ell^{1+k/2}}{ \ln \ell} \to \infty,
\end{equation}
thus it does not follow that the left-hand side of \eqref{logESanti}
converges to $-\infty$.

Note that the proof (and theorem) does hold in the case $k=2$, due to the extra
condition $r \gg 1$. It is also correct in the general case if one adds the
condition that $r \geq const.$ in the limit,
or if one replaces the condition $nr = k \ell \ln \ell $ with $nr = k \ell \ln n$.
(The proof can easily be adapted in the latter case.)

The authors also presented a second, weaker result for the upper 
bounds \cite[Theorem~4]{SBV:range_assignment},
using a different proof technique. (The result is also reported within 
Theorem~3 in \cite{SanBlo:transmitting_range}.) They claim the following:

\begin{quotepar}
   ``Suppose $n$ nodes are placed in $[0,\ell]$ according to the uniform distribution.
   If $rn \in \Omega(\ell \log \ell)$, then the $r$-homogeneous
   range assignment is a.a.s. connecting.''
\end{quotepar}

(Here $rn \in \Omega(\ell \log \ell)$ means that $\ell \log \ell = O(rn)$, 
the $r$-homogeneous range assignment refers to the system considered above, 
and ``a.a.s. connecting'' means $P_\conndb \to 1$ in our notation.)
This statement conflicts with our results as well, with a similar 
counterexample as above (where $k$ is chosen sufficiently large). 
In fact, it is also in conflict with \cite[Theorem 5]{SanBlo:transmitting_range}.
The proof, as given by the authors, relies on
Theorem~2 in \cite{SBV:range_assignment}, which reads:

\begin{quotepar}
   ``Assume $n$ nodes are displaced at random in $[0,\ell]$. 
   Then, the probability that the $r$-homogeneous range assignment
   is connecting is at least \\$1-(\ell-r) (1-\frac{r}{\ell})^{n}$.''
\end{quotepar}

To see that this result is incorrect, remember that $P_\conndb$ does not
change when scaling both $r$ and $\ell$ together, i.e. when replacing
$\ell$ with $\lambda \ell$ and $r$ with $\lambda r$, where $\lambda > 0$
is arbitrary. Exploiting this property, the above theorem leads to the
conclusion that for any fixed $\ell$ and $r$,
\begin{equation}
  \forall \lambda > 0: \;
  P_\conndb \geq 1 - \lambda (\ell-r) (1-\frac{r}{\ell})^{n};
\end{equation}
however, this would obviously result in $P_\conndb = 1$
for all parameter values.

The root cause of this error seems to be in the proof of the named theorem:
Here, the authors define certain events $\event{DISCONNECTED}_\ell^{s,r}$,
where $s \in [0,\ell-r]$ is a continuous parameter, such that
\begin{equation} \label{disconnUnion}
   \event{DISCONNECTED}_\ell = \bigcup_{s \in [0,\ell-r]} \event{DISCONNECTED}_\ell^{s,r};
\end{equation}   
$\event{DISCONNECTED}_\ell$ is the complement of our event $\conndb$.
They then argue as follows.

\begin{quotepar}
 ``An upper bound to $P(\event{DISCONNECTED}_\ell)$ can be derived 
 by summing the probabilities $P(\event{DISCONNECTED}_\ell^{s,r})$
 for all possible values of $s$. We thus have:
\begin{equation*}
  P(\event{DISCONNECTED}_\ell) \leq \int_{0}^{\ell-r} P(\event{DISCONNECTED}_\ell^{s,r}) ds
  \quad \text{ [\ldots].'' }
\end{equation*}
\end{quotepar}
However, unlike the analogue case for a finite union of events,
this is not a valid consequence of Eq.~\eqref{disconnUnion} --
passing to the integral for the ``summation of probabilities'' 
is by no means justified.

For the lower bounds on the probability of connectedness, 
Theorem~5 in \cite{SanBlo:transmitting_range} states:

\begin{quotepar}
``Assume that $n$ nodes, each with transmitting range $r$, are distributed
uniformly and independently at random in $R = [0,\ell]$, and assume that
$rn = (1-\epsilon) \ell \ln \ell$ for some $0 < \epsilon < 1$.
If $ r = r(\ell) \in \Theta(\ell^\epsilon)$, then the communication graph
is not connected w.h.p.''
\end{quotepar}

Here ``not connected w.h.p.'' corresponds to 
$P_\conndb \not\to 1$ in our notation. This theorem \emph{is} compatible 
with the present work. In fact, using that the relation between $\ln \ell$
and $\ln n$ is fixed by the requirement $r \in \Theta(\ell^\epsilon)$, 
one can use Theorem~\ref{pconnDbAsympThm} to show that 
$P_\conndb \to 0$ under the conditions given.

Under more general conditions, Theorem 6 in \cite{SanBlo:transmitting_range} 
claims the following result:

\begin{quotepar}
``Assume that $n$ nodes, each with transmitting range $r$, are distributed
uniformly and independently at random in $R = [0,\ell]$ and assume that
$r = r(\ell) \ll \ell$ and $n = n(\ell) \gg 1$.
If $ rn \ll \ell \ln \ell$, then the communication graph
is not connected w.h.p.''
\end{quotepar}
 
This statement again is incompatible with the results in Sec.~\ref{disconnectedSec}.
As a counterexample, let $n = \ln \ell$ and $r = \ell / \ln\ln \ell$, thus 
fulfilling all prerequisites of the theorem. In this case, we have
\begin{equation}
  n r = \frac{\ell \; \ln \ell}{ \ln\ln\ell} 
  = \frac{\ln \ell}{ (\ln\ln\ell)^2 } \; \ell \; \ln n 
  \geq 2 \ell \; \ln n \quad
  \text{for large $\ell$;}
\end{equation}
so Corollary~\ref{SantiCompareCorr} tells us that $P_\conndb \to 1$.

For its proof, the cited Theorem 6 of \cite{SanBlo:transmitting_range} 
relies on \cite[Theorem~4]{SanBlo:connectivity}. The proof of that theorem,
located in the Appendix of \cite{SanBlo:connectivity}, is in fact 
inconclusive: Defining $C := \ell/r$, the authors note

\begin{quotepar}
``Observe that the condition $\ell \ll rn \ll \ell \log \ell$ implies that
$C \ll n \ll C \log C$ [\ldots].''
\end{quotepar}

However, in the general case, this implication does not hold: 
In the above counterexample, we have in fact
\begin{equation}
  \ell \ll rn = \frac{\ell\; \ln \ell}{ \ln \ln \ell} \ll \ell \ln \ell,
\end{equation}
but it follows from $C = \ell/r = \ln \ln \ell$ that 
\begin{equation}
  n = \ln \ell \not\ll  \ln \ln \ell \;   \ln \ln \ln \ell = C \ln C.
\end{equation}
Thus, one cannot conclude $n \ll C \log C$, and
the subsequent arguments in \cite{SanBlo:connectivity}
do not apply.

In conclusion, let us briefly mention that Theorem~7 of \cite{SanBlo:transmitting_range},
which summarizes most of the propositions discussed above, 
does consequently not hold in the stated form.

\backmatter

%Index of Notation
\chapswitch

\specialChapter{Index of Notation}   \label{NotationApp}
%\documentclass{article}

%\usepackage{german}
%\usepackage[latin1]{inputenc}

%\usepackage[all]{xy}

%\begin{document}

%\cleardoublepage

\thispagestyle{plain}

%\pdfdest name{notation} xyz

\paragraph{Asymptotic behaviour of functions.}
For two functions $f,g$, we write $f = O(g)$ or $f(x) = O(g(x))$
if $f(x) \leq g(x) \cdot \mathrm{const}$ in the limit being considered 
(usually $x \to \infty$). 
The notation $f = \Theta(g)$ is used as an abbreviation for
$f = O(g) \wedge g = O(f)$.
We write $f \sim g$ to denote that $f(x)/g(x) \to 1$.
For sequences rather than functions, we use similar notation.
The sign ``$\approx$'' is used in a more qualitative sense in heuristic 
argumentation, meaning ``approximately equal to'' (in a sense to be
specified later).

\paragraph{Vector notation.}
Vectors (i.e. elements of some $\rbb^n$) are denoted by 
boldface symbols, while their components are denoted in normal typeface;
e.g.: $\xv = (x_1,\ldots,x_n)$. We do not always explicitly specify
the dimension of the underlying vector space where it is apparent 
from the context.

\paragraph{Symbols and abbreviations.}
The following table lists symbols and abbreviations
frequently used in the text, with a reference to their
definition or first occurrence.

\bigskip 

\begin{center}
\begin{tabular}{lp{0.55\textwidth}r}
symbol \quad
&
description
&
reference
\\
\hline
$\conndb$
&
event of connected MANET with disconnected boundary conditions
&
Eq.~\eqref{sortedIntConn}
\\
$\connpb$
&
event of connected MANET with periodic boundary conditions
&
Eq.~\eqref{connpbDef}
\\
$\discdb{k}$
&
event of $k$-disconnected MANET with disconnected boundary conditions
&
Eq.~\eqref{discpbDef}
\\
$\discpb{k}$
&
event of $k$-disconnected MANET with periodic boundary conditions
&
Eq.~\eqref{discdbDef}
\\
$\unitv{j}$
&
$j$-th standard unit vector in $\rbb^n$
\\
$\expect{F}$
&
expectation value of a random variable $F$
&
Eq.~\eqref{expectDef}
\\
$M_\event{EV}$
&
subset of $\Omega_n$ associated with an event $\event{EV}$
&
Sec.~\ref{RandomVarSec}
\\
$\ell$
&
spatial extent of the MANET
&
Eq.~\eqref{OmegaNOneD}
\\
$n$
&
total number of network nodes in the MANET
&
Sec.~\ref{ProbModelSec}
\\
$\nbb$
&
$= \{ 1,2,3,\ldots \}$
\\
$\nbb_0$
&
$= \{ 0,1,2,\ldots \}$
\\
$P_\event{EV}$
&
probability of an event $\event{EV}$
&
Eq.~\eqref{probabDef}
\\
$O(f) $
&
see ``asymptotic behaviour of functions'' above
\\
\end{tabular}
\end{center}

\begin{flushright}
(continued on next page)
\end{flushright}
\clearpage
\medskip
\begin{center}
\begin{tabular}{lp{0.55\textwidth}r}
symbol {\qquad}
&
description
&
reference
\\
\hline
$r$
&
radio range of a MANET node
&
Sec.~\ref{modeldefSec}
\\
$\rbb^+$
&
$= \{x \in \rbb \,|\, x > 0\}$
\\
$\rbb^+_0$
&
$= \{x \in \rbb \,|\, x \geq 0\}$
\\
$T_n$
&
top surface of $V_n$
&
Eq.~\eqref{simplexSurfaceDef}
\\
$V_n$
&
$n$-dimensional standard simplex
&
Eq.~\eqref{simplexDef}
\\
\\
%
%
% griechische Buchstaben
%
%
$\delta(\cdotarg) d^nx$
&
delta-valued integration measure
&
Eq.~\eqref{surfaceIntDelta}
\\
$\chi_\event{EV}$
&
characteristic function of an event $\event{EV}$
&
Eq.~\eqref{charactDef}
\\
$\eta$
&
limit of $n \rho - \ln n$ as $n \to \infty$
&
Thm.~\ref{pconnPbAsympThm}
\\
$\muEqualT{n}$
&
measure of equal distribution on $T_n$
&
Prop.~\ref{tnEqualProp}
\\
$\nu$
&
limit of $n \rho$ as $n \to \infty$
&
Def.~\ref{intensiveDefn}
\\
$\rho$
&
$=r/\ell$, normalized radio range
&
Sec.~\ref{modeldefSec}
\\
$\theta(\cdotarg)$
&
Heaviside's theta function
&
Eq.~\eqref{thetaDef}
\\
$\Theta(f) $
&
see ``asymptotic behaviour of functions'' above
\\
$\Omega_n$
&
sample space for a MANET with $n$ nodes
&
Eq.~\eqref{OmegaDef}
\\
\\
%
%
% andere Symbole
%
%
$\overline{F}$
&
$= \expect{F}$, expectation value of a random variable $F$ 
&
Eq.~\eqref{expectDef}
\\
$[x]$
&
$= \max \{ k \in \zbb \, | \, k \leq x \}$, Gauss bracket of $x$
&
\\
$M^c$
&
complement of a set $M$
\\
\end{tabular}
\end{center}

%\end{document}

%List of References

\renewcommand{\bibname}{References}

\specialChapterSilent{References}{
\bibliographystyle{myalpha}
\bibliography{manet}
}

%Lists of Tables and Figures - on two facing pages

\specialChapterSilentLR{\listtablename}{\listoftables}
\specialChapterSilentLR{\listfigurename}{\listoffigures}

\clearpage
\pagestyle{plain}

%\cleardoublepage

\section*{Acknowledgments}

I would like to thank Jörg Roth for supervising this Master's thesis,
as well as for supplying the detail data lying on the
base of his article cited as \cite{Rot:critical_mass}.

\end{document}